\documentclass[aps,pre,reprint,nofootinbib]{revtex4-1} 

\usepackage{amsthm}
\usepackage{amssymb}   
\usepackage{mathtools} 
\usepackage{hyperref}
\hypersetup{colorlinks=true,linkcolor=blue,urlcolor=blue,citecolor=blue}
\usepackage{accents}
\usepackage{tensor}
\usepackage[cal=boondox]{mathalfa}
\usepackage{lipsum}

\usepackage[T1]{fontenc}

\usepackage{mathtools}

\begin{document}
\title{Geometry of the parameter space of a quantum system: Classical point of view}
\author{Javier Alvarez-Jimenez}
\email{javier171188@hotmail.com}

\author{Diego Gonzalez}
\email{diego.gonzalez@correo.nucleares.unam.mx}

\author{Daniel Guti\'errez-Ruiz}
\email{daniel.gutierrez@correo.nucleares.unam.mx}

\author{J. David Vergara}
\email{vergara@nucleares.unam.mx}

\affiliation{Departamento de F\'isica de Altas Energ\'ias, Instituto de Ciencias Nucleares, Universidad Nacional Aut\'onoma de M\'exico, Apartado Postal 70-543, Ciudad de M\'exico, 04510, M\'exico}

\date{\today}

\begin{abstract}
The local geometry of the parameter space of a quantum system is described by the quantum metric tensor and the Berry curvature, which  are two fundamental objects that play a crucial role in understanding geometrical aspects of condensed matter physics. We consider classical integrable systems and report a new approach to obtain the classical analogs of the quantum metric tensor and the Berry curvature. An advantage of this approach is that it can be applied to a wide variety of classical systems corresponding to quantum systems with bosonic and fermionic degrees of freedom. Our approach arises from the semiclassical approximation of the Berry curvature and the quantum metric tensor in the Lagrangian formalism.  We also exploit this semiclassical approximation to establish, for the first time, the relation between the quantum metric tensor and its classical counterpart. We illustrate and validate our approach by applying it to five systems: the generalized harmonic oscillator, the symmetric and linearly coupled harmonic oscillators, the singular Euclidean oscillator, and a spin-half particle in a magnetic field. Finally, we mention some potential applications of this approach and possible generalizations that can be of interest in the field of condensed matter physics.	
\end{abstract}

\maketitle

\section{Introduction}

The Berry curvature~\cite{Berry45} has become one of the fundamental concepts in modern physics. It is a geometrical property of quantum states, that gives rise to a geometric (Berry) phase when the system undergoes an adiabatic evolution along a closed path in the parameter space. The Berry curvature, which was initially used to study the anomalous Hall conductivity~\cite{Jungwirth2002,Haldane2004}, has recently acquired a growing interest in the field of condensed matter~\cite{XiaoNiu2010,Armitage2018,Rostami2018}. For instance, in crystalline solids, where the parameters are the crystal momentum, the Berry curvature is well recognized as an intrinsic property of the band structure~\cite{XiaoNiu2010}. Another fundamental concept, which is related to the Berry curvature, is the quantum metric tensor~\cite{Provost1980}. This metric is also a geometrical property of quantum states and provides a measure of the distance, on the parameter space,  between two quantum states with infinitesimally different parameters. Among the full range of applications, the quantum metric tensor is used for detecting the presence of quantum phase transitions~\cite{SHI-JIAN2010,VenutiZanardi2007,Zanardi2007,Kumar2012,Zhu2006} and also plays an essential role to extract topological charges of tensor monopoles in condensed matter physics~\cite{Palumbo2018,Palumbo2019}. Although the quantum metric tensor and the Berry curvature are two completely different objects, they turn out to be part of the same geometrical structure: the quantum geometric tensor. Specifically, its real part is the quantum metric tensor, and its imaginary part corresponds to the Berry curvature.

Recently, in the context of the holographic principle, it has been proposed~\cite{Miyaji2015} (see also \cite{BAK2016200,Trivella_2017}) a Lagrangian approach to compute the quantum metric tensor through Euclidean path integrals. This idea was taken further in Ref.~\cite{Alvarez-Jimenez2017}, where it was shown that such an approach can be extended to obtain the quantum geometric tensor.  In contrast to the standard Hamiltonian approach, the Lagrangian approach involves perturbations of the Lagrangian in the parameter space rather than perturbations of quantum states, which makes it suitable not only for quantum field theories but also for systems where the exact solution is not available~\cite{AlvarezVergara-2019}.

On the other hand, in the context of thermodynamic systems, the classical counterparts of the quantum metric tensor were introduced in the seminal works of Weinhold~\cite{Weinhold1975} and Ruppeiner \cite{Ruppeiner1979}, and were subsequently implemented in the broader frame known as geometrothermodynamics~\cite{Quevedo2007}. In the framework of classical integrable systems (those for which it is possible to introduce action-angle variables), both the Berry curvature and the quantum metric tensor have a counterpart. The classical analog of the Berry curvature is the curvature of the Hannay connection\cite{Hannay1985}, which brings about an extra (Hannay) angle picked up by the angle variables when the system performs a closed adiabatic loop in the parameter space. The semiclassical relation between the Berry phase and the Hannay angle was given in Ref.~\cite{Berry1985} and verified in a variety of systems (see Refs.~\cite{Biswas1990,Brihaye1993}, for instance).   An alternative derivation of such a semiclassical relation by using coherent states was provided in Ref.~\cite{Maamache1990}. In the case of the quantum metric tensor, its classical analog was recently introduced in Ref.~\cite{GGVmetric2019} and corresponds to a classical metric in the parameter space that measures the distance between two points in phase space with infinitesimally different parameters. The essential element involved in the classical metric is the generating function of the canonical transformation that infinitesimally shifts the phase space variables with respect to the parameters of the Hamiltonian. An important property of this classical metric is that it produces the same (or almost the same) parameter structure of the quantum metric tensor, modulo a Bohr-Sommerfeld quantization rule for action variables ~\cite{GGVmetric2019} .

In this paper, we consider classical integrable systems and propose a new approach to compute the classical analogs of the quantum metric tensor and the Berry curvature, which relies on the Lagrangian formalism. We achieve this by performing the semiclassical approximation of the quantum geometric tensor in the Lagrangian approach, which allows us to obtain new expressions for the classical metric introduced in Ref.~\cite{GGVmetric2019} and the curvature of the Hannay connection. The  new expressions involve averages over classical variables (the angle variables), and it is in this context that we use the term "classical" throughout this work. An advantage of our approach is that it does not require the generating function considered in Ref.~\cite{GGVmetric2019} and, therefore, can be applied to a wide range of systems, including those where such a function is unknown. Furthermore, we extend our classical approach to consider fermionic systems, by including classical Grassmannian variables~\cite{Gozzi2388_1987,Gozzi2752_1990}. In this paper, we also point out the semiclassical relation between the quantum and classical metrics. This result,  together with that of Berry in Ref.~\cite{Berry1985}, completes the set of semiclassical relations between the quantum structures and their corresponding classical analogs in the framework of classical integrable systems. We illustrate and corroborate our approach and the semiclassical relation in five examples, obtaining  the expected results.

The structure of the paper is the following. In Sec.~\ref{sec:LagrangianA} we describe the Lagrangian approach proposed in  Ref.~\cite{Alvarez-Jimenez2017} to calculate the quantum metric tensor and the Berry curvature from the quantum geometric tensor. In Sec.~\ref{sec:Classical} we carry out the semiclassical approximation to the quantum geometric tensor to obtain the classical counterparts of the quantum metric tensor and the Berry curvature. In Sec.~\ref{Examples} we apply our approach to compute the classical metric and curvature in five systems. The first system is the archetypal generalized harmonic oscillator, for which the classical metric and curvature are well known~\cite{GGVmetric2019}. The second one is a system of symmetric coupled harmonic oscillators, which has been used in the study of quantum entanglement~\cite{Bombelli1986,Srednicki1993,Chandran2019} and circuit complexity~\cite{Chapman2018,Jefferson2017}. The third system considered is that of linearly coupled harmonic oscillators, which can also be found in the study of quantum entanglement~\cite{Kim_2005,Paz2008,Makarov2018}. The next system analyzed is the singular Euclidean oscillator, which plays the role of the confinement potential in quantum ring models~\cite{Bellucci2011}. The final system studied is a particle with  spin $1/2$ in a magnetic field, which has served as a prototypical example to illustrate the appearance of the Berry phase~\cite{Berry45} and the Hannay angle~\cite{Gozzi2388_1987,Gozzi2752_1990}.  In this example, we extend our approach to fermionic systems using classical Grassmannian variables. In each model, we compare the classical results with the quantum counterparts. To conclude, Sec.~\ref{Conclusions} presents final remarks and explores some new directions for further work. 


\section{Quantum metric tensor and Berry curvature: Lagrangian formalism}\label{sec:LagrangianA}

In this section, we recall some basic aspects on the quantum metric tensor and the Berry curvature in the Lagrangian formalism. Here, we follow the description introduced in Ref.~\cite{Alvarez-Jimenez2017}.  Consider a quantum system that during the Euclidean time interval $\tau\in(-\infty,0)$ is represented by a path integral with the Lagrangian $L(q(\tau),\dot{q}(\tau);x)$, where $q(\tau)=\{q^a(\tau)\}$  ($a,b,\dots\! =1,\dots,n$) are the configuration variables and $x=\{x^i\}$ ($i,j,\dots\!=1,\dots,N$) is a set of $N\geq 1$ parameters, that parametrize an $N$-dimensional parameter space $\mathcal{M}$. Let us now assume that at time $\tau=0$, the system is subject to an adiabatic perturbation arising from a small change of the parameters $x \rightarrow x'=x+\delta x$, so that for the Euclidean time interval $\tau\in(0,\infty)$, the system is described by the perturbed Lagrangian $L'=L-{\cal O}_{i}{\rm d}x^{i}$, where 
\begin{eqnarray}\label{qua:operatorO}
\hspace{-2mm}{\cal O}_{i}(\tau)\! :=\! {\cal O}_{i}(q(\tau),\dot{q}(\tau);x)\! =\!\!-\!\left(\partial_iL(q(\tau),\dot{q}(\tau);x)\right)_{q(\!\tau\!),\dot{q}(\!\tau\!)}
\end{eqnarray}
are the Lagrangian deformations. Throughout the paper, we use the notation $\partial_i := \partial/ \partial x^i$.

Let $|\psi_{0}\rangle$ and $|\psi'_{0}\rangle$ be the ground states for the original and deformed systems, respectively. Then the fidelity is defined by ${\cal F}(x,x+{\rm d}x)\equiv|\langle \psi'_{0},\tau\rightarrow\infty|\psi_{0},\tau\rightarrow-\infty\rangle|$ and gives a measure of the change of the system by turning on the deformations. In the path-integral formalism, the series expansion of the fidelity leads to ${\cal F}(x,x+{\rm d}x)\simeq 1-\frac{1}{2}G_{ij}^{(0)}(x){\rm d}x^{i}{\rm d}x^{j}$, where
\begin{align}\label{geomtesor}
G_{ij}^{(0)}(x)=\frac{1}{\hbar^{2}} \intop_{-\infty}^{0}{\rm d}\tau_{1}\intop_{0}^{\infty}{\rm d}\tau_{2} \bigg(& \langle\hat{{\cal O}}_{i}(\tau_{1})\hat{{\cal O}}_{j}(\tau_{2})\rangle_{0}\nonumber\\
& - \langle\hat{{\cal O}}_{i}(\tau_{1})\rangle_{0} \langle\hat{{\cal O}}_{j}(\tau_{2})\rangle_{0} \bigg)\, ,
\end{align}
is the quantum geometric tensor. Here, $\langle \cdot \rangle_{0}$ denotes the (functional) expectation value with respect to the undeformed Lagrangian $L$, namely 
\begin{equation}
\langle \hat{{\cal A}} \rangle_{0} = \frac{1}{Z_0} \int \mathcal{D}q \left[ \exp\left( -\frac{1}{\hbar} \int_{-\infty}^{\infty} {\rm d}\tau L \right) {\cal A}(q) \right]\, ,
\end{equation}
where $Z_0$ is a normalization. It must be noticed that the undeformed system already contains all the ``interactions'' associated  to the parameters $x$.  

Assuming time-reversal symmetry for the two-point functions, i.e.,
$\langle\hat{{\cal O}}_{i}(-\tau_{1})\hat{{\cal O}}_{j}(-\tau_{2})\rangle_{0}=\langle\hat{{\cal O}}_{i}(\tau_{1})\hat{{\cal O}}_{j}(\tau_{2})\rangle_{0}$, the quantum metric tensor $g_{ij}^{(0)}(x)$ arises from the real part of Eq.~(\ref{geomtesor}):
\begin{align}\label{qua:metric}
g_{ij}^{(0)}(x) =\frac{1}{\hbar^{2}} \intop_{-\infty}^{0}  {\rm d}\tau_{1} \intop_{0}^{\infty}  {\rm d}\tau_{2}\Bigg(&\frac{1}{2}\langle[\hat{{\cal O}}_{i}(\tau_{1}),\hat{{\cal O}}_{j}(\tau_{2})]_{+}\rangle_{0} \nonumber\\
&-\langle\hat{{\cal O}}_{i}(\tau_{1})\rangle_{0}\langle\hat{{\cal O}}_{j}(\tau_{2})\rangle_{0}\Bigg) \, ,
\end{align}
whereas the Berry curvature $F_{ij}^{(0)}(x)$ corresponds to the imaginary part:
\begin{equation}\label{qua:curvature}
F_{ij}^{(0)}(x)=-\frac{1}{i\hbar^{2}}\intop_{-\infty}^{0}{\rm d}\tau_{1}\intop_{0}^{\infty}{\rm d}\tau_{2}\langle[\hat{{\cal O}}_{i}(\tau_{1}),\hat{{\cal O}}_{j}(\tau_{2})]_{-}\rangle_{0}\, ,
\end{equation}
where $[\cdot,\cdot]_{+}$ and $[\cdot,\cdot]_{-}$ stand for the anticommutator and the commutator, respectively.

\section{Classical counterparts of the quantum metric tensor and the Berry curvature}\label{sec:Classical}

The aim of this section is to derive the classical counterparts of the quantum metric tensor~(\ref{qua:metric}) and the Berry curvature~(\ref{qua:curvature}) for classical integrable systems. 
Before presenting the derivation, it is convenient to  perform a Wick rotation to real time, $\tau \rightarrow i t$, and write the operators $\hat{{\cal O}}_{i}(t)$ in terms of the phase space operators, $\hat{q}=\{\hat{q}^a\}$ and $\hat{p}=\{\hat{p}_a\}$, as 
\begin{equation}\label{qua:operatorO2}
\hat{{\cal O}}_{i}(t)=\hat{{\cal O}}_{i}(\hat{q}(t),\hat{p}(t);x)=\left(\partial_i\hat{H}(\hat{q}(t),\hat{p}(t);x)\right)_{\hat{q}(t),\hat{p}(t)}\, ,
\end{equation}
where $\hat{H}(\hat{q}(t),\hat{p}(t);x)$ is the Hamiltonian of the system for the real time interval $t\in(-\infty,0)$. Notice that we are working in the Heisenberg representation. By integrating the Heisenberg equations for  the operators $\hat{q}(t)$ and $\hat{p}(t)$ with the initial conditions $\hat{q}(t=0)=\hat{q}_0$ and $\hat{p}(t=0)=\hat{p}_0$ where $\hat{q}_0$ and $\hat{p}_0$ are the usual position and momentum operators (those in the Schr\"odinger representation), we can express $\hat{q}(t)$ and $\hat{p}(t)$ in terms of  $\hat{q}_0$, $\hat{p}_0$ and time. This allows to define the operator
 \begin{equation}\label{qua:operatorO3}
\hat{\lambda}_{i}(t)\!:=\hat{\lambda}_{i}(t,\hat{q}_0,\hat{p}_0;x)\!=\!\hat{{\cal O}}_{i}(\hat{q}(t,\hat{q}_0,\hat{p}_0;x),\hat{p}(t,\hat{q}_0,\hat{p}_0;x);x)\, ,
 \end{equation}
and write the expectation values involved in Eqs.~(\ref{qua:metric}) and~(\ref{qua:curvature}) in the following alternative way:
\begin{subequations}
	\begin{align}
		\langle\hat{{\cal O}}_i(t)\rangle_0 & = \langle \psi_0(x)|   \hat{\lambda}_i(t)| \psi_0(x) \rangle \nonumber\\
		&=\int {\rm d}q_0 \, \psi_{0}^*(q_0;x) \, \lambda_i \left(t,q_0,-i\hbar \frac{\partial}{\partial q_0};x \right) \, \psi_{0}(q_0;x)\, , \label{qua:exp1}
	\end{align}
	\begin{align}
		&\langle[ \hat{{\cal O}}_i(t_1),\hat{{\cal O}}_j(t_2)]_{\pm} \rangle_{0} \nonumber \\
		&= \langle \psi_0(x) |  [ \hat{\lambda}_i(t_1),\hat{\lambda}_j(t_2) ]_{\pm}| \psi_0(x) \rangle \nonumber \\
		&= \int \! {\rm d}q_0 \, \psi_0^*(q_0;x)  \nonumber \\
		& \times\! \! \left[\! \lambda_i \! \left(\! t_1,q_0,-i\hbar \frac{\partial}{\partial q_0};\! x \right)\! ,\lambda_j\! \left(\! t_2,q_0,-i\hbar \frac{\partial}{\partial q_0};\! x \right) \right]_{\pm}\! \! \psi_{0}(q_0;x) ,\label{qua:exp2}
	\end{align}
\end{subequations}
where $\int dq_0 \equiv\prod_{a=1}^{n} \int dq^a_0$ and $\psi_{0}(q_0;x)\equiv\langle q_0|\psi_{0}(x)\rangle $ is the (normalized) wave function.

We begin the derivation by considering the  semiclassical approximation to the wave function $\psi_{0}(q_0;x)$ \cite{Maslov1981,Brack1997}:
\begin{equation}\label{semiclassicalwave}
\psi_0(q_0;x)= \sum_\alpha a_{(\alpha)}(q_0,I_0;x) {\rm e}^{ \frac{i}{\hbar}S^{(\alpha)}(q_0,I_0;x)}\, ,
\end{equation}
with 
\begin{equation}
\left(a_{(\alpha)}(q_0,I;x)\right)^2 = \frac{1}{(2 \pi)^n } \det\left(\frac{\partial \varphi_0^{(\alpha)a}}{\partial q_0^b}\right)\, .
\end{equation}
Here,  $S^{(\alpha)}(q_0,I_0;x)$ is the multivalued  generating function of the canonical transformation from $(q_0,p_0)$ to the (initial) action-angle variables  $(\varphi^{(\alpha)}_0,I_0)$ with $\varphi^{(\alpha)}_0=\{\varphi^{(\alpha)a}_0\}$ and $I_0\equiv I(t)=\{I_a\}$, and $\alpha$ labels different branches of $S^{(\alpha)}(q_0,I_0;x)$. Plugging the wave function~(\ref{semiclassicalwave}) into Eqs.~(\ref{qua:exp1}) and (\ref{qua:exp2}), and bearing in mind that terms corresponding to different branches $\alpha$ exhibit a strongly oscillating behavior, we have, respectively,
\begin{subequations}
	\begin{align}
	&\langle\hat{{\cal O}}_i \! (t)\rangle_0\! =\!\int \! \frac{{\rm d} q_0}{(2 \pi)^n}  \sum_\alpha  \det \! \left(\frac{\partial \varphi_0^{(\alpha)a}}{\partial q_0^b}\right) \lambda_{i}\! \left(t,q_0,\frac{\partial S^{(\alpha)}}{\partial q_0};x \right) \!+\!O(\hbar) \, , \label{qua:exp3}
    \end{align}
	\begin{align}
&	\langle[ \hat{{\cal O}}_i(t_1),\hat{{\cal O}}_j(t_2)]_{\pm} \rangle_{0} \nonumber\\
	& =\int \frac{{\rm d}q_0}{(2 \pi)^n}  \sum_\alpha  \det\left(\frac{\partial \varphi_0^{(\alpha)a}}{\partial q_0^b}\right) \nonumber\\
	& \times\left[ \lambda_i \left(t_1,q_0, \frac{\partial S^{(\alpha)}}{\partial q_0};x \right),\lambda_j\left(t_2,q_0, \frac{\partial S^{(\alpha)}}{\partial q_0};x \right) \right]_{\pm}+O_{\pm}(\hbar) \, ,\label{qua:exp4}
	\end{align}
\end{subequations}
where $O(\hbar)$ and $O_{\pm}(\hbar)$ denote terms at least linear in $\hbar$. 

In the case of bosonic operators, we arrive at the classical limit by replacing $\frac{\partial S^{(\alpha)}}{\partial q_0}$ by $p^{(\alpha)}_0$, the anticommutators $[\, ,]_{+}$ by simple products, and the commutators $[\, ,]_{-}$ by the non-equal-time Poisson brackets  $i \hbar \{\, ,\}_{(q_0,p^{(\alpha)}_0)}$, 
\begin{eqnarray}
&&\{f(t_1),g(t_2)\}_{(q_0,p^{(\alpha)}_0)}=\nonumber \\ && \sum_{a=1}^{n} \left( \frac{\partial f(t_1)}{\partial q^a_0} \frac{\partial g(t_2)}{\partial p^{(\alpha)}_{a0}} - \frac{\partial f(t_1)}{\partial p^{(\alpha)}_{a0}} \frac{\partial g(t_2)}{\partial  q^a_0}\right)\, . 
\end{eqnarray}
To achieve the case with fermionic operators, we have to replace the anticommutators by the corresponding Poisson brackets and the commutators by appropriate products. Restricting our derivation to the case of bosonic operators and treating $O(\hbar)$ and $O_{\pm}(\hbar)$ as small compared to the remaining terms in Eqs.~(\ref{qua:exp3}) and~(\ref{qua:exp4}), the expectation values become
\begin{subequations}
	\begin{align}
	&\langle\hat{{\cal O}}_i(t)\rangle_0 \approx \int  \frac{{\rm d}q_0}{(2 \pi)^n}  \sum_\alpha  \det\left(\frac{\partial \phi_0^{(\alpha)a}}{\partial q_0^b}\right) \lambda_{i} (t,q_0, p^{(\alpha)}_0;x)\, , \label{qua:exp5}
    \end{align}
    \begin{align}
	\langle[ \hat{{\cal O}}_i(t_1),\hat{{\cal O}}_j(t_2)]_{+} \rangle_{0}  \approx & 2 \int  \frac{{\rm d}q_0}{(2 \pi)^n}  \sum_\alpha  \det\left(\frac{\partial \phi_0^{(\alpha)a}}{\partial q_0^b}\right) \nonumber\\
	& \times \lambda_i(t_1,q_0, p^{(\alpha)}_0;x ) \lambda_j(t_2,q_0,p^{(\alpha)}_0;x ) \, ,\label{qua:exp6}
    \end{align}
    \begin{align}
	\langle[ \hat{{\cal O}}_i&(t_1),\hat{{\cal O}}_j(t_2)]_{-} \rangle_{0} \nonumber \\
	&\approx i \hbar \!\int \! \frac{{\rm d}q_0}{(2 \pi)^n}  \sum_\alpha  \det\left(\frac{\partial \phi_0^{(\alpha)a}}{\partial q_0^b}\right) \nonumber\\
	& \ \ \ \times \{ \lambda_i(t_1,q_0, p^{(\alpha)}_0;x),\lambda_j (t_2,q_0,p^{(\alpha)}_0;x) \}_{(q_0,p^{(\alpha)}_0)}\, .\label{qua:exp7}
	\end{align}
\end{subequations}

Next, we make $\varphi^{(\alpha)}_0$ single-valued
by choosing the branch defined by $0\leq \varphi_0 <  2\pi$. This in turn makes $p^{(\alpha)}_0$ single--valued, in which case the index $\alpha$ can be omitted. Moreover, performing the change of variables $q_0 \rightarrow \varphi_0$,  we have
\begin{subequations}
	\begin{align}
	\langle\hat{{\cal O}}_i(t)\rangle_0\approx\frac{1}{(2 \pi)^n} \oint {\rm d}\varphi_0 \,  \lambda_{i}(t) =\langle \lambda_{i} (t) \rangle\, , \label{qua:exp8}
    \end{align}
	\begin{align}
	\langle[ \hat{{\cal O}}_i(t_1),\hat{{\cal O}}_j(t_2)]_{+} \rangle_{0} & \approx  \frac{2}{(2 \pi)^n} \oint {\rm d}\varphi_0  \lambda_i(t_1) \lambda_j(t_2)\nonumber \\
	&= 2 \langle \lambda_i\left(t_1\right) \lambda_j(t_2) \rangle \, , \label{qua:exp9}
	\end{align}
	\begin{align}
	\langle[ \hat{{\cal O}}_i(t_1),\hat{{\cal O}}_j(t_2)]_{-} \rangle_{0} & \approx\frac{i \hbar}{(2 \pi)^n} \oint {\rm d}\varphi_0 \{ \lambda_i(t_1),\lambda_j(t_2) \}_{(q_0,p_0)}\nonumber \\
	&  = i \hbar \langle \{ \lambda_i\left(t_1\right),\lambda_j(t_2) \}_{(q_0,p_0)} \rangle \, , \label{qua:exp10}
	\end{align}
\end{subequations}
where $\lambda_{i}(t):=\lambda_{i}(t,q_0,p_0;x)$ and $\langle f \rangle=\frac{1}{(2 \pi)^{n}}\oint {\rm d}\varphi_0 f $, with $\oint d\varphi_0 \equiv\prod_{a=1}^{n} \int_{0}^{2 \pi}d\varphi^a_0$, is the average of $f(\varphi_0,I;x)$ over the angle variables $\varphi_0$.  

At this point, it is not difficult to realize that the classical functions  $\lambda_{i}(t,q_0,p_0;x)$ are given by
\begin{equation}\label{class:functionO3}
\lambda_{i}(t,q_0,p_0;x)={\cal O}_{i}(q(t,q_0,p_0;x),p(t,q_0,p_0;x);x)\, ,
\end{equation}
where
\begin{equation}\label{class:funcO}
{\cal O}_{i}(t)={\cal O}_{i}(q(t),p(t);x)\!=\!\left(\partial_iH(q(t),p(t);x)\right)_{q(t),p(t)}\, ,
\end{equation}
with $H(q(t),p(t);x)$ being the classical analog of the operator $\hat{H}(\hat{q}(t),\hat{p}(t);x)$, and the classical variables $q(t)$ and $p(t)$ have been expressed in terms of $q_0=q(t=0)$, $p_0=p(t=0)$ and time by solving the Hamilton equations of motion. Clearly, an alternative expression for ${\cal O}_{i}(t)$ in terms of $q(t)$ and $\dot{q}(t)$ is given by Eq.~(\ref{qua:operatorO}).

Substituting Eqs.~(\ref{qua:exp8}) and~(\ref{qua:exp9}) into Eq.~(\ref{qua:metric}) (after a Wick rotation $\tau\rightarrow i t$), we arrive at
\begin{equation}\label{relationmetrics}
g_{ij}^{(0)}(x) \approx\frac{1}{\hbar^{2}} g_{ij}(I;x)\, ,
\end{equation}
where 
\begin{equation}
g_{ij}(I;x)\!=\!-\!\intop_{-\infty}^{0}{\rm d}t_{1}\intop_{0}^{\infty} {\rm d} t_{2}\,\left(\langle\lambda_{i}(t_{1})\lambda_{j}(t_{2})\rangle-\langle\lambda{}_{i}(t_{1})\rangle\langle\lambda_{j}(t_{2})\rangle\right)\, ,\label{class:metric}
\end{equation}
is the \textit{classical analog of the quantum metric tensor}~(\ref{qua:metric}). Since we have considered classical integrable systems, this implies that Eq.~(\ref{class:metric}) is an alternative expression of the classical metric proposed in Ref.~\cite{GGVmetric2019}, and as such provides a measure of the distance, on the parameter space, between two points in phase space corresponding to infinitesimally different parameters. The advantage of the metric~(\ref{class:metric}) over the metric introduced in Ref.~\cite{GGVmetric2019} is that it does not require for its calculation the knowledge of the generating functions $G_i(q,I;x) := - ( \partial_i S^{(\alpha)} )_{q,I}$, which are not always easy to determine. Notice that Eq.~(\ref{relationmetrics}) provides the semiclassical relation between the quantum metric tensor and the classical metric, and its validity is limited to the case of classical integrable systems.  Furthermore, it should be pointed out that the relation~(\ref{relationmetrics}) does not hold in the presence of quantum anomalies that may result from the particular form of the operators $\hat{\lambda}_{i}(t)$, which in turn depend on the Hamiltonian. We will see this through examples. One way to anticipate the presence of such quantum anomalies would be the appearance of loop diagrams in the computation of Eq.~(\ref{qua:metric})~\cite{AlvarezVergara-2019}. Nevertheless, this issue requires further investigation and is beyond the scope of the present study.

Inserting Eq.~(\ref{qua:exp10}) into Eq.~(\ref{qua:curvature}) (after a Wick rotation), we obtain
\begin{equation}\label{relationcurvatures}
F_{ij}^{(0)}(x) \approx\frac{1}{\hbar} F_{ij}(I;x) \, ,
\end{equation}
where
\begin{equation}
F_{ij}(I;x)=\intop_{-\infty}^{0}{\rm d}t_{1}\intop_{0}^{\infty} {\rm d} t_{2}\, \langle\{\lambda_{i}(t_{1}),\lambda_{j}(t_{2})\}_{(q_0,p_0)}\rangle\, ,\label{class:curv}
\end{equation}
is the \textit{classical analog of the Berry curvature}~(\ref{qua:curvature}). This entails that  Eq.~(\ref{class:curv}) is actually the curvature of Hannay's connection, and can be used to calculate the Hannay angle. On the other hand, Eq.~(\ref{relationcurvatures}) is precisely  the  semiclassical relation between the Berry curvature and Hannay's curvature established by Berry in Ref.~\cite{Berry1985}. 

 Note that while relation~(\ref{relationcurvatures}) for the curvatures involves the factor $1/\hbar$, the analogous relation~(\ref{relationmetrics}) for the metrics  involves the factor $1/\hbar^{2}$. The origin of these different factors can be traced 	back to the fact that in the former case the replacement of the commutators  by the Poisson brackets introduces $\hbar$, while in the latter case  the corresponding replacement of the anticommutators does not.

Finally, it can be checked that Eqs.~(\ref{class:metric}) and (\ref{class:curv}) are invariant under the (gauge) canonical transformation
\begin{equation}\label{gclas:gauge}
\varphi'^a_0 = \varphi^a_0 + \frac{\partial \lambda(I;x)}{\partial I_a}, \qquad I'_a=I_a,
\end{equation}
where $\lambda(I;x)$ is an arbitrary function of the actions $I$ and the parameters $x$. Note that, in the quantum setting, this is analogous to the invariance of the quantum metric tensor and the Berry curvature under a phase transformation.

\section{Illustrative  examples}\label{Examples}

\subsection{Generalized harmonic oscillator}

It is instructive to illustrate and corroborate the results of the previous section in a well-known example. Let us take the generalized harmonic oscillator, which is described by the classical Hamiltonian 
\begin{equation}\label{gho:classicalH}
H=\frac{1}{2}\left(X q^2+2Yqp+Zp^2\right)\, ,
\end{equation}
or by the Lagrangian
\begin{equation}\label{gho:classicalL}
L=\frac{1}{2Z}\left(\dot{q}^2-2Yq \dot{q}-\omega^2 q^2\right)\, ,
\end{equation}
where $x=\{x^{i}\}=(X,Y,Z)$ ($i,j,\dots\!=1,2,3$) are the adiabatic parameters and $\omega:=(XZ-Y^{2})^{1/2}$ is the angular frequency of the system. The corresponding functions ${\cal O}_i(t)$ obtained from either Eq.~(\ref{gho:classicalH}) or Eq.~(\ref{gho:classicalL}) are
\begin{subequations}
	\begin{align}
		&{\cal O}_{1}(t) =\frac{1}{2} q^{2}\, , \label{gho:lambda1}\\
		&{\cal O}_{2}(t)  = \frac{q}{Z}\left(\dot{q}-Yq\right) =q p\, , \label{gho:lambda2}\\
		&{\cal O}_{3}(t) = \frac{1}{2Z} \left(\dot{q}-Yq\right)^2  =\frac{1}{2} p^{2}\, ,\label{gho:lambda3}
	\end{align}
\end{subequations}
where $p=\frac{1}{Z}(\dot{q}-Yq)$. Next, we need to express these functions in terms of time $t$ and the initial conditions $q_{0}=q(t=0)$ and $p_{0}=p(t=0)$. With this in mind, we write the solution for the variables $(q,p)$ as follows:
\begin{subequations}
	\begin{align}
		&q(t)=q_{0} \cos\omega t + \frac{1}{\omega} \left(Z p_0 + Y q_0\right) \sin\omega t\, , \label{gho:qt}\\
		&p(t)=p_{0} \cos\omega t- \frac{1}{\omega}\left(Y p_0 + X q_0\right)  \sin\omega t\, .\label{gho:pt}
	\end{align}
\end{subequations}
Then, substituting (\ref{gho:qt}) and (\ref{gho:pt}) into (\ref{gho:lambda1})--(\ref{gho:lambda3}), we arrive at the functions $\lambda_{i}(t)$:
\begin{subequations}
	\begin{align}
		\lambda_{1}(t) =&\frac{1}{2} \left[ q_{0} \cos\omega t + \frac{1}{\omega} \left(Z p_0 + Y q_0\right) \sin\omega t\right]^2\, , \label{gho:lambda1a}\\
		\lambda_{2}(t)  =& q_0 p_0 \cos^2 \omega t -\frac{(X q_0^2-Z p_0^2)}{\omega}  \cos\omega t \sin\omega t  \nonumber\\
		&- \frac{(X q_0 + Yp_0) (Y q_0+Z p_0)}{\omega^2} \sin^2 \omega t\, ,  \\
		\lambda_{3}(t) =& \frac{1}{2} \left[p_{0} \cos\omega t- \frac{1}{\omega}\left(Y p_0 + X q_0\right)  \sin\omega t\right]^{2}\, .\label{gho:lambda3a}
	\end{align}
\end{subequations}
Furthermore, the initial conditions $(q_0,p_0)$ in terms of the action-angle variables $(\varphi_0,I)$ read
\begin{subequations}
	\begin{align}
		&q_0=\left(\frac{2ZI}{\omega}\right)^{1/2}\sin\varphi_0\, ,\label{gho:q}\\
		&p_0=\left( \frac{2ZI}{\omega} \right)^{1/2} \left(-\frac{Y}{Z}\sin\varphi_0+\frac{\omega}{Z}\cos\varphi_0\right)\, .\label{gho:p}
	\end{align}
\end{subequations}

Having obtained the functions $\lambda_{i}(t)$,  we proceed to calculate the classical metric~(\ref{class:metric}). Substituting Eqs.~(\ref{gho:q}) and (\ref{gho:p}) into Eqs.~(\ref{gho:lambda1a})--(\ref{gho:lambda3a}) and defining $\Lambda_{ij}:=\langle\lambda_{i}(t_{1})\lambda_{j}(t_{2})\rangle -\langle\lambda_{i}(t_{1})\rangle \langle\lambda_{j}(t_2)\rangle$ where $\langle f\rangle=\frac{1}{2\pi}\intop_{0}^{2\pi}{\rm d}\varphi_0 f$, the averages  involved in Eq.~(\ref{class:metric}) lead to
\begin{subequations}
	\begin{align}
		&\Lambda_{11} =\frac{Z^2 I^{2}}{8\omega^{2}}\cos 2\omega t_{12}\, , \label{gho:L11}\\
		&\Lambda_{12} =-\frac{Z I^{2}}{4\omega^{2}} \left( Y \cos 2\omega t_{12} -\omega \sin 2\omega t_{12} \right) \, , \\
		&\Lambda_{13} =-\frac{I^{2}}{8\omega^{2}} \left[ (XZ-2Y^2) \cos 2\omega t_{12} +2Y\omega \sin 2\omega t_{12} \right]\, ,\\
		&\Lambda_{22} =\frac{XZ I^{2}}{2\omega^{2}}\cos 2\omega t_{12}\, , \\
		&\Lambda_{23} = -\frac{X I^{2}}{4\omega^{2}} \left( Y \cos 2\omega t_{12} -\omega \sin 2\omega t_{12} \right)\, ,\\
		&\Lambda_{33} =\frac{X^2 I^{2}}{8\omega^{2}}\cos 2\omega t_{12}\, ,\label{gho:L33}
	\end{align}
\end{subequations}
where $t_{12}=t_1-t_2$.  Inserting Eqs.~(\ref{gho:L11})--(\ref{gho:L33}) into Eq.~(\ref{class:metric}) and using
\begin{subequations}
	\begin{align}
		&\intop_{-\infty}^{0}{\rm d}t_{1}\intop_{0}^{\infty} {\rm d} t_{2} \cos 2\omega t_{12} = -\frac{1}{4\omega^2}\, , \label{gho:avgcos}\\
		&\intop_{-\infty}^{0}{\rm d}t_{1}\intop_{0}^{\infty} {\rm d} t_{2} \sin 2\omega t_{12} = 0\, ,\label{gho:avgsin}
	\end{align}
\end{subequations}
we find the classical metric $g_{ij}(I;x)$:
\begin{eqnarray}\label{gho:classmetric}
g_{ij}(I;x)=\frac{I^{2}}{32\omega^{4}}\left(\begin{array}{ccc}
Z^{2} & -2YZ & 2Y^{2}-XZ\\
-2YZ & 4XZ & -2XY\\
2Y^{2}-XZ & -2XY & X^{2}
\end{array}\right)\, , \nonumber \\
\end{eqnarray}	
which is exactly the same as that obtained in Ref.~\cite{GGVmetric2019} by using the classical metric that involves generating functions. This corroborates our claim above that Eq.~(\ref{class:metric}) yields the same results of the metric introduced in Ref.~\cite{GGVmetric2019}.  We can also compare Eq.~(\ref{gho:classmetric}) with the quantum metric tensor for the ground state obtained by using Eq.~(\ref{qua:metric}) (we refer the reader to Ref~\cite{Alvarez-Jimenez2017} where the computations are done):
\begin{eqnarray}\label{gho:QIM}
g^{(0)}_{ij}(x)=\frac{1}{32\omega^{4}}\left(\begin{array}{ccc}
Z^{2} & -2YZ & 2Y^{2}-XZ\\
-2YZ & 4XZ & -2XY\\
2Y^{2}-XZ & -2XY & X^{2}
\end{array}\right)\, . \nonumber \\
\end{eqnarray}
Thus, both metrics (\ref{gho:classmetric}) and  (\ref{gho:QIM})  have the same parameter structure. Furthermore, by using the identification $I^{2}=\hbar^2$, it is readily seen that they satisfy the relation~(\ref{relationmetrics}).

We now turn to the calculation of the curvature~(\ref{class:curv}). It follows from Eqs.~(\ref{gho:lambda1a})--(\ref{gho:lambda3a}) that the average of the non-equal-time Poisson brackets among the functions $\lambda_{i}(t)$ with respect to the initial conditions $(q_{0},p_{0})$ are
\begin{subequations}
	\begin{align}
		\langle\{\lambda_{1}(t_{1}),\lambda_{2}(t_{2})\}_{(q_{0},p_{0})}\rangle=&\frac{Z I}{\omega^{2}} \left( \omega \cos 2\omega t_{12} + Y \sin 2\omega t_{12} \right)\, ,\\
		\langle\{\lambda_{1}(t_{1}),\lambda_{3}(t_{2})\}_{(q_{0},p_{0})}\rangle=&-\frac{I}{2\omega^{2}} [2 Y \omega \cos 2\omega t_{12} \nonumber\\
		&+ (2Y^2-XZ) \sin 2\omega t_{12} ]\, ,\\
		\langle\{\lambda_{2}(t_{1}),\lambda_{3}(t_{2})\}_{(q_{0},p_{0})}\rangle=&\frac{X I}{\omega^{2}} \left( \omega \cos 2\omega t_{12} + Y \sin 2\omega t_{12} \right)\, .
	\end{align}
\end{subequations}	
Then, plugging these expressions into Eq.~(\ref{class:curv}) and using Eqs.~(\ref{gho:avgcos}) and (\ref{gho:avgsin}), we arrive at the components of the classical curvature
\begin{eqnarray}\label{gho:classcurvature}
F_{12}(I;x)=-\frac{ZI}{4\omega^{3}}\, ,\ \
F_{13}(I;x)=\frac{YI}{4\omega^{3}}\, , \ \
F_{23}(I;x)=-\frac{XI}{4\omega^{3}}. \nonumber \\
\end{eqnarray}
The components (\ref{gho:classcurvature}) are precisely those found in Ref.~\cite{GGVmetric2019} via the use of generating functions (see also Ref.~\cite{Berry1985}). Furthermore, in the quantum case the components of the Berry curvature for the ground state coming from Eq.~(\ref{qua:curvature}) are (see Ref~\cite{Alvarez-Jimenez2017}):
\begin{eqnarray}\label{gho:Bcurvature}
F_{12}^{(0)}(x)=-\frac{Z}{8\omega^{3}}\, ,\ \
F_{13}^{(0)}(x)=\frac{Y}{8\omega^{3}}\, ,\ \ F_{23}^{(0)}(x)=-\frac{X}{8\omega^{3}}\, . \nonumber \\
\end{eqnarray}
By comparing~(\ref{gho:classcurvature}) and (\ref{gho:Bcurvature}), and taking into account the Bohr-Sommerfeld quantization rule for action variable $I=\hbar/2$, it is direct to check that these curvatures satisfy the relation (\ref{relationcurvatures}).

Hence, in this example, besides showing the applicability of Eqs.~(\ref{class:metric}) and~(\ref{class:curv}), we have verified that these expressions yield the expected results for the classical metric and curvature.

A final remark is that to make contact with the quantum results, we have seen that different powers of the action variable must be quantized differently. Actually, this feature will appear in all our illustrative examples. The reason is that the computation of the expectation values in the quantum metric naturally incorporates loop diagrams ~\cite{AlvarezVergara-2019}, and since that is a purely quantum effect, we must postulate a different quantization rule for superior powers of $I$. In any case, if this empirical quantization rule were not taken into account, the quantum and classical metrics would only differ by a global numeric factor, although their parameter dependence would be identical.


\subsection{Symmetric coupled harmonic oscillators}

We now want to illustrate our approach by applying it to a system with two degrees of freedom. Since many physical models are based on coupled oscillators, we  have chosen as our second example the system consisting of two coupled harmonic oscillators described by the Hamiltonian
\begin{equation}\label{sco:Hamil}
H=\frac{1}{2} \left[ p_{1}^{2}+p_{2}^{2}+k (q_{1}^{2}+q_{2}^{2})+k^{\prime} (q_{1}-q_{2})^{2}  \right]\, ,
\end{equation}
where $x=\{x^{i}\}=(k,k^{\prime})$ ($i,j,\dots\!=1,2$) are the adiabatic parameters. In particular, this system has been widely used to clarify the physical basis of quantum entanglement~\cite{Bombelli1986,Srednicki1993,PandoZayas2015,Chandran2019} and gain some intuition towards circuit complexity~\cite{Jefferson2017}.


From Eq.~(\ref{sco:Hamil}), it is clear that the associated functions ${\cal O}_{i}(t)$ are
\begin{equation}\label{sco:lambda}
{\cal O}_{1}(t) =\frac{1}{2}(q_{1}^{2}+q_{2}^{2})\, , \qquad {\cal O}_{2}(t)=\frac{1}{2}(q_{1}-q_{2})^{2}\, .
\end{equation}
For our purposes, it is convenient to introduce the transformation
\begin{subequations}
	\begin{align}
		&Q_{1}=\frac{1}{\sqrt{2}}(q_{1}+q_{2})\, , \qquad Q_{2}=\frac{1}{\sqrt{2}}(q_{1}-q_{2})\, ,\label{sco:transfq}\\
		&P_{1}=\frac{1}{\sqrt{2}}(p_{1}+p_{2})\, , \qquad P_{2}=\frac{1}{\sqrt{2}}(p_{1}-p_{2})\, ,
	\end{align}
\end{subequations}
which allows us to write the Hamiltonian~(\ref{sco:Hamil}) as $H=\frac{1}{2}( P_{1}^{2} +P_{2}^{2}  + \omega_{1}^{2} Q_{1}^{2} + \omega_{2}^{2} Q_{2}^{2} )$ where 
\begin{equation}
\omega_{1}^{2}=k \, , \qquad \omega_{2}^{2}=k+2k^{\prime}\, , 
\end{equation}
are the frequencies of the uncoupled oscillators. The new coordinates as functions of time read
\begin{eqnarray}\label{sco:Qt}
Q_a(t)=Q_{a0} \cos\omega_a t + \frac{P_{a0}}{\omega_a} \sin\omega_a t\, , \qquad (a=1,2)\, ,
\end{eqnarray}
where $Q_{a0}=Q_{a}(t=0)$ and $P_{a0}=P_a(t=0)$ are the initial conditions. With this
at hand, the functions $\lambda_{i}(t)$, in terms of $Q_{0}=\{Q_{a0}\}$, $P_{0}=\{P_{a0}\}$ and time, turn out to be
\begin{subequations}
	\begin{align}
		\lambda_{1}(t)  =&\frac{1}{2}\bigg[\frac{P_{10}\left(P_{10}\sin^{2}\omega_{1}t+\omega_{1}Q_{10}\sin2\omega_{1}t\right)}{\omega_{1}^{2}} \nonumber\\&+\frac{P_{20}\left(P_{20}\sin^{2}\omega_{2}t+\omega_{2}Q_{20}\sin2\omega_{2}t\right)}{\omega_{2}^{2}}\nonumber \\
		&+Q_{10}^{2}\cos^{2}\omega_{1}t+Q_{20}^{2}\cos^{2}\omega_{2}t\bigg]\, , \label{sco:lambda2a}\\
		\lambda_{2}(t)  =&\frac{(P_{20}\sin\omega_{2}t+\omega_{2}Q_{20}\cos\omega_{2}t)^{2}}{\omega_{2}^{2}}\, . \label{sco:lambda2b}
	\end{align}
\end{subequations}
Furthermore, since the system is separable in the new coordinates, we can easily write the initial conditions in terms of the action-angle variables $(\varphi_0,I)$:
\begin{eqnarray}\label{sco:Q0P0}
\hspace{-4mm}Q_{a0}=\left(\frac{2 I_a}{\omega_a}\right)^{1/2} \! \sin\varphi_{a0}\,  \ \ P_{a0}=\left(2 \omega_a I_a\right)^{1/2} \cos\varphi_{a0}\, .
\end{eqnarray}

We are now ready to compute the classical metric~(\ref{class:metric}). By plugging Eq.~(\ref{sco:Q0P0}) into Eqs.~(\ref{sco:lambda2a}) and (\ref{sco:lambda2b}) and defining $\Lambda_{ij}:=\langle\lambda_{i}(t_{1})\lambda_{j}(t_{2})\rangle -\langle\lambda_{i}(t_{1})\rangle \langle\lambda_{j}(t_2)\rangle$ where $\langle f\rangle=\frac{1}{(2\pi)^2}\intop_{0}^{2\pi}{\rm d }\varphi_{10}\intop_{0}^{2\pi}{\rm d }\varphi_{20}\,f$,  the averages appearing in Eq.~(\ref{class:metric}) become
\begin{subequations}
	\begin{align}
		&\Lambda_{11} =\frac{I_{1}^{2}}{8\omega_{1}^{2}}\cos2\omega_{1} t_{12}+\frac{I_{2}^{2}}{8\omega_{2}^{2}}\cos2\omega_{2} t_{12}\, , \\
		&\Lambda_{12} =\frac{I_{2}^{2}}{4\omega_{2}^{2}}\cos2\omega_{2}t_{12}\, ,\\
		&\Lambda_{22} =\frac{I_{2}^{2}}{2\omega_{2}^{2}}\cos2\omega_{2}t_{12}\, ,
	\end{align}
\end{subequations}
with $t_{12}=t_1-t_2$. Inserting these expressions into  Eq.~(\ref{class:metric}) and using Eq.~(\ref{gho:avgcos}) (with $\omega_a$ instead of $\omega$), the classical metric tensor $g_{ij}(I;x)$  turns out to be
\begin{equation}\label{sco:metric}
g_{ij}(I;x)=\frac{1}{32}\begin{pmatrix}
\frac{I_{1}^{2}}{\omega_{1}^{4}}+\frac{I_{2}^{2}}{\omega_{2}^{4}} & \frac{2 I_{2}^{2}}{\omega_{2}^{4}}\\
\frac{2 I_{2}^{2}}{\omega_{2}^{4}} & \frac{4 I_{2}^{2}}{\omega_{2}^{4}}
\end{pmatrix}\, ,
\end{equation}
and its determinant is $\det [g_{ij}(I;x)]=I_{1}^{2}I_{2}^{2}/256\omega_{1}^{4}\omega_{2}^{4}\neq0$. Notice that this metric can be expressed, in compact form, as the sum of two terms corresponding to the two uncoupled oscillators:
\begin{equation}\label{sco:metric2}
g_{ij}(I,x)=\frac{\partial_i \omega_1 \partial_j \omega_1}{8\omega_1^2} I_1^2 + \frac{\partial_i \omega_2 \partial_j \omega_2}{8\omega_2^2} I_2^2\, .
\end{equation}
 This decomposition is a feature of the system under consideration and is not always possible for entangled systems, as we will see in the example given in the next subsection. On the other hand, we remark that each action variable $I_a$ in Eq.~(\ref{sco:metric2}) is associated with an independent mode of the system, not a particular (coupled) oscillator. This means, since each normal mode involves both coupled oscillators, that the classical metric~(\ref{sco:metric2}) cannot not be factored into single-particle terms, and hence it may capture some properties of entanglement.

At this point, it is instructive to compare Eq.~(\ref{sco:metric}) with the quantum metric tensor obtained from Eq.~(\ref{qua:metric}). With this in mind, let us consider the following Hamiltonian operator
\begin{equation}\label{sco:qHamil}
\hat{H}=\frac{1}{2} \left[ \hat{p}_{1}^{2}+\hat{p}_{2}^{2}+k (\hat{q}_{1}^{2}+\hat{q}_{2}^{2})+k^{\prime} (\hat{q}_{1}-\hat{q}_{2})^{2}  \right]\, ,
\end{equation}
for which the corresponding operators $\hat{{\cal O}}_i(t)$ are
\begin{equation}\label{sco:qO}
\hat{{\cal O}}_{1}(t) =\frac{1}{2}(\hat{q}_{1}^{2}+\hat{q}_{2}^{2})\, , \qquad \hat{{\cal O}}_{2}(t)=\frac{1}{2}(\hat{q}_{1}-\hat{q}_{2})^{2}\, .
\end{equation}
Using a transformation analogous to~(\ref{sco:transfq}), these operators take the form
\begin{equation}\label{sco:qO2}
\hat{{\cal O}}_{1}(t) =\frac{1}{2}(\hat{Q}_{1}^{2}+\hat{Q}_{2}^{2}), \qquad \hat{{\cal O}}_{2}(t)=\hat{Q}_{2}^{2}\, ,
\end{equation}
where $\hat{Q}_a(t)$ in terms of the annihilation and creation operators, $\hat{b}_a(t)$ and $\hat{b}_a^\dagger(t)$, read
\begin{equation}\label{sco:operatorQ}
\hat{Q}_a(t) = \sqrt{\frac{\hbar}{2 \omega_a}} \left( \hat{b}_a(t) + \hat{b}_a^\dagger(t) \right)\qquad (a=1,2)\, ,
\end{equation}
with the frequencies $\omega_{1}^{2}=k$ and $\omega_{2}^{2}=k+2k^{\prime}$.

Defining $\beta_{ij}:=\langle \hat{{\cal O}}_i(t_1) \hat{{\cal O}}_j(t_2) \rangle_0 \!-\! \langle \hat{{\cal O}}_i(t_1)\rangle_0 \langle \hat{{\cal O}}_j(t_2)\rangle_0$ and using Eq.~(\ref{sco:qO2}) together with Eq.~(\ref{sco:operatorQ}), the expectation values in Eq.~(\ref{qua:metric}) turn out to be
\begin{subequations}
	\begin{align}
	&\beta_{11} = \frac{\hbar^2}{8\omega_1^2} {\rm e}^{-2 i \omega_1 t_{12}} + \frac{\hbar^2}{8\omega_2^2}{\rm e}^{-2 i \omega_2 t_{12}} \, , \\
	&\beta_{12} = \frac{\hbar^2}{4 \omega_2^2} {\rm e}^{-2 i \omega_2 t_{12}} \, ,\\
	&\beta_{22} = \frac{\hbar^2}{2\omega_2^2} {\rm e}^{-2i \omega_2 t_{12}}\, .
	\end{align}
\end{subequations}
Thus, plugging these results into Eq.~(\ref{qua:metric}) and integrating, we find the quantum metric tensor
\begin{equation}\label{sco:qmetric}
g^{(0)}_{ij}(x)=\frac{1}{32}\begin{pmatrix}
\frac{1}{\omega_{1}^{4}}+\frac{1}{\omega_{2}^{4}} & \frac{2 }{\omega_{2}^{4}}\\
\frac{2}{\omega_{2}^{4}} & \frac{4}{\omega_{2}^{4}}
\end{pmatrix}\, ,
\end{equation}
which has a nonvanishing determinant, $\det[g^{(0)}_{ij}(x)]=1/256\omega_{1}^{4}\omega_{2}^{4}$.

By comparing Eqs.~(\ref{sco:metric}) and (\ref{sco:qmetric}), it is straightforward to see that both metrics have exactly the same dependence on the parameters of the system. Furthermore, with the identifications $I_1^2=I_2^2=\hbar^2$, it is not hard to verify that they satisfy the relation~({\ref{relationmetrics}}).

Let us now compute the classical curvature~(\ref{class:curv}). From Eqs.~(\ref{sco:lambda2a}) and (\ref{sco:lambda2b}), the average over the angle variables of the non-equal-time Poisson Brackets among  $\lambda_{1}(t_{1})$ and $\lambda_{2}(t_{2})$ with respect to $(Q_{0},P_{0})$ gives \begin{equation}
\langle\{\lambda_{1}(t_{1}),\lambda_{2}(t_{2})\}_{(Q_{0},P_{0})}\rangle=-\frac{I_{2}}{\omega_{2}^{2}}\sin2\omega_{2}t_{12}\, ,
\end{equation}
which, together with (\ref{gho:avgsin}) (with $\omega_2$ instead of $\omega$), implies that the classical curvature (\ref{class:curv}) vanishes, $F_{12}(I;x)=0$. 

In the quantum case, the quantum geometric tensor  turns out to be purely real, from which it follows that the  Berry curvature~(\ref{qua:curvature}) vanishes, $F_{12}^{(0)}(x)=0$. Therefore, the classical and quantum curvatures lead to the same result.

We see from this example that all the information contained in the parameter space of the quantum system is encoded in the parameter space of the associated classical system. This is a non-trivial result and suggests, motivated by the geometric characterization  of quantum entanglement by using the Fubini-Study metric~\cite{Aniello_Marmo_2010,Volkert_Marmo_2010}, that the classical metric might also carry information about  entanglement.  This contrasts with the common idea that entanglement is a  quantum phenomenon that is absent in classical systems.


\subsection{Linearly coupled harmonic oscillators}

We now turn to an example where the semiclassical relation (\ref{relationmetrics}) is not satisfied. Consider the case of two coupled harmonic oscillators described by the Hamiltonian 
\begin{equation}\label{lco:classicalH}
H=\frac{1}{2}\left(p_1^2+p_2^2+A q_1^2 + B q_2^2 + C q_1 q_2 \right)\, ,
\end{equation}
where $x=\{x^{i}\}=(A,B,C)$ ($i,j,\dots\!=1,2,3$) are the adiabatic parameters, which are assumed to satisfy $A\neq B$. This system has also been extensively studied in the context of quantum entanglement~\cite{Kim_2005,Paz2008,Makarov2018}. For instance, it has been shown that for certain parameters it exhibits a very large quantum entanglement~\cite{Makarov2018}. It is worth noting that the  Hamiltonian~(\ref{sco:Hamil}) is not a particular case of Eq.~(\ref{lco:classicalH}), and hence the results of the preceding action cannot be obtained from the results presented below.

From (\ref{lco:classicalH}), it is straightforward to see that the functions ${\cal O}_{i}(t)$ are
\begin{equation}\label{lco:lambda1}
{\cal O}_1(t)=\frac12 q_1^2\, , \quad {\cal O}_2(t)=\frac12  q_2^2\, , \quad {\cal O}_3(t)=\frac12 q_1 q_2\, .
\end{equation}  
To better deal with these functions, let us consider the transformation 
\begin{subequations}
	\begin{align}
		&(Q_1,Q_2)=(q_1 \cos\alpha  - q_2\sin \alpha,q_1 \sin\alpha  + q_2\cos \alpha)\, ,\label{lco:T1}\\
		&(P_1,P_2)=(p_1 \cos\alpha  - p_2\sin \alpha,p_1 \sin\alpha  + p_2\cos \alpha)\, ,\label{lco:T2}
	\end{align}
\end{subequations}
where $\tan\alpha = \frac{\epsilon}{|\epsilon|} \sqrt{\epsilon^2+1} - \epsilon$ with $\epsilon=\frac{B-A}{C}$. Note that $\tan\alpha \in (-1,1)$, and then $\alpha \in (-\pi/4,\pi/4)$. Using this transformation, the Hamiltonian~(\ref{lco:classicalH}) is diagonalized as $H=\frac{1}{2}(P_1^2+P_2^2+\omega_1^2 Q_1^2 + \omega_2^2 Q_2^2)$ where $\omega_1^2=A-\frac{C}{2}\tan\alpha$ and $\omega_2^2=B+\frac{C}{2}\tan\alpha$ are the angular frequencies of the uncoupled harmonic oscillators. Furthermore, the functions ${\cal O}_i$ in (\ref{lco:lambda1}) take the form
\begin{eqnarray}\label{lco:lambda2}
{\cal O}_i(Q,P;x)= \omega_1 Q_1^2 \partial_i \omega_1 +  \omega_2 Q_2^2 \partial_i \omega_2 +  (\omega_2^2-\omega_1^2) Q_1 Q_2 \partial_i \alpha\, .\nonumber \\
\end{eqnarray} 
Now, for the new coordinates we have
\begin{eqnarray}\label{lco:Qt}
Q_a(t)=Q_{a0} \cos\omega_a t + \frac{P_{a0}}{\omega_a} \sin\omega_a t\, , \quad (a=1,2)\, ,
\end{eqnarray}
where the initial conditions $Q_{a0}=Q_{a}(t=0)$ and $P_{a0}=P_a(t=0)$ in terms of the action-angle variables $(\varphi_0,I)$ are
\begin{eqnarray}\label{lco:Q0P0}
\hspace{-4mm}Q_{a0}=\left(\frac{2 I_a}{\omega_a}\right)^{1/2}\!\! \sin\varphi_{a0}\, ,  \ \ P_{a0}=\left(2 \omega_a I_a\right)^{1/2} \cos\varphi_{a0}\, .
\end{eqnarray}
We can readily obtain the functions $\lambda_i(t)$ in terms of $Q_{0}=\{Q_{a0}\}$, $P_{0}=\{P_{a0}\}$ and time by plugging Eq.~(\ref{lco:Qt}) into Eq.~(\ref{lco:lambda2}), so we do not write them here.

We now proceed to calculate the classical metric~(\ref{class:metric}). Using $\langle f\rangle=\frac{1}{(2\pi)^2}\intop_{0}^{2\pi}{\rm d}\varphi_{10}\intop_{0}^{2\pi}{\rm d}\varphi_{20}\,f$ and the resulting functions $\lambda_i(t)$, the averages that appear in Eq.~(\ref{class:metric}) are
\begin{align}\label{lco:avg}
\langle\lambda_{i}&(t_{1})\lambda_{j}(t_{2})\rangle-\langle\lambda{}_{i}(t_{1})\rangle\langle\lambda_{j}(t_{2})\rangle \nonumber \\
 &= \frac{1}{2} \partial_i \omega_1 \partial_j \omega_1 I_1 \cos 2\omega_1t_{12} +\frac{1}{2} \partial_i \omega_2 \partial_j \omega_2 I_2 \cos 2\omega_2 t_{12} \nonumber \\
&\ \ \ +\partial_i\alpha \partial_j\alpha \left(\omega_1^2-\omega_2^2\right)^2 \frac{I_1 I_2}{\omega_1 \omega_2} \cos\omega_1 t_{12} \, \cos\omega_2 t_{12}\, ,
\end{align}
where $t_{12}=t_1-t_2$. Substituting (\ref{lco:avg}) into (\ref{class:metric}) and taking into account (\ref{gho:avgcos}) (with $\omega_a$ instead of $\omega$) together with
\begin{equation}
\intop_{-\infty}^{0}{\rm d} t_{1}\intop_{0}^{\infty} {\rm d} t_{2}  \cos\omega_1 t_{12} \, \cos\omega_2 t_{12} = -\frac{\omega_1^2+\omega_2^2}{(\omega_1^2-\omega_2^2)^2}\, ,
\end{equation}
we obtain the classical metric
\begin{align}\label{lco:classmetric}
g_{ij}(I,x)=&\frac{\partial_i \omega_1 \partial_j \omega_1}{8\omega_1^2} I_1^2 + \frac{\partial_i \omega_2 \partial_j \omega_2}{8\omega_2^2} I_2^2 \nonumber \\
& + \partial_i\alpha \partial_j\alpha \left(\frac{\omega_1}{\omega_2} 
+  \frac{\omega_2}{\omega_1} \right) I_1 I_2\, . 
\end{align}
Notice that, as compared with Eq.~(\ref{sco:metric2}), this metric has the extra term proportional to the product $I_1 I_2$ and, because of that, it cannot be separated in terms corresponding to the two uncoupled oscillators. Explicitly, the metric~(\ref{lco:classmetric}) reads
\begin{eqnarray}
g_{ij}(I,x)\!=\!\frac{1}{32} \! \left[\frac{I_1^2}{ \omega_1^4} M_{ij}\! + \! \frac{I_2^2}{\omega_2^4} N_{ij} \! + \! \left(\!\frac{\omega_1}{\omega_2} \! + \! \frac{\omega_2}{\omega_1} \! \right) \frac{8 I_1 I_2}{(\omega_2^2-\omega_1^2)^2} L_{ij} \right], \nonumber \\
\end{eqnarray}
where
\begin{equation}
M_{ij}=\!\frac{1}{4}
\begin{pmatrix}
(1+\mu)^2 & \nu ^2 & -(1+\mu)\nu \\
\nu ^2 & (1-\mu )^2 & -  (1-\mu )\nu \\
- (1+\mu)\nu & -(1-\mu )\nu & \nu ^2
\end{pmatrix}\, ,
\end{equation}
\begin{equation}
N_{ij}=\frac{1}{4} 
\begin{pmatrix}
(1-\mu )^2 & \nu ^2 & (1-\mu )\nu \\
\nu ^2 & (1+\mu)^2 & (1+\mu)\nu \\
(1-\mu )\nu & (1+\mu) \nu & \nu ^2
\end{pmatrix}\, ,
\end{equation}
\begin{equation}
L_{ij}=
\begin{pmatrix}
\nu ^2 & -\nu ^2 & \nu  \mu  \\
-\nu ^2 & \nu ^2 & -\nu  \mu  \\
\nu  \mu  & -\nu  \mu  & \mu ^2
\end{pmatrix}\, ,
\end{equation}
with $\mu:=\cos2\alpha=\frac{\epsilon}{\sqrt{\epsilon^2+1}}$ and $\nu:=\sin2\alpha=\frac{1}{\sqrt{\epsilon^2+1}}$. It is worth noting that this metric has a nonvanishing determinant, $\det [g_{ij}(I;x)]= (\omega_1^2+\omega_2^2)I_1^3 I_2^3 /4096 (\omega_1^2-\omega_2^2)^2 \omega_1^5 \omega_2^5 \neq 0$.

It is interesting to contrast the classical metric (\ref{lco:classmetric}) with that coming from the quantum metric tensor~(\ref{qua:metric}). In the quantum setting, the Hamiltonian operator of the system reads
\begin{equation}\label{lco:quH}
\hat{H}=\frac{1}{2}\left(\hat{p}_1^2+\hat{p}_2^2+A \hat{q}_1^2 + B \hat{q}_2^2 + C \hat{q}_1 \hat{q}_2 \right)\, ,
\end{equation}
and the quantum operators $\hat{{\cal O}}_i(t)$ are
\begin{eqnarray}\label{lco:quanO1}
\hat{{\cal O}}_1(t)=\frac12 \hat{q}_1^2, \quad \hat{{\cal O}}_2(t)=\frac12 \hat{q}_2^2, \quad\hat{{\cal O}}_3(t)=\frac12 \hat{q}_1 \hat{q}_2\, .
\end{eqnarray}  
By following a procedure analogous to that used for Eq.~(\ref{lco:lambda2}), these operators can be written as
\begin{eqnarray}\label{lco:operatorO}
\hat{{\cal O}}_i(t) = \omega_1 \hat{Q}_1^2 \partial_i\omega_1 +\omega_2 \hat{Q}_2^2 \partial_i \omega_2-\left(\omega_1^2-\omega_2^2 \right)\hat{Q}_1 \hat{Q}_2 \partial_i \alpha\, ,\nonumber \\
\end{eqnarray}
where the operators $\hat{Q}_a(t)$ in terms of the creation and annihilation operators are given by Eq.~(\ref{sco:operatorQ}), but with the frequencies $\omega_1^2=A-\frac{C}{2}\tan\alpha$ and $\omega_2^2=B+\frac{C}{2}\tan\alpha$.

 Using Eq.~(\ref{lco:operatorO}) together with Eq.~(\ref{sco:operatorQ}), we compute the expectation values appearing in Eq.~(\ref{qua:metric}), obtaining
\begin{align}
&\langle \hat{{\cal O}}_i(t_1) \hat{{\cal O}}_j(t_2) \rangle_0 - \langle \hat{{\cal O}}_i(t_1)\rangle_0 \langle \hat{{\cal O}}_j(t_2)\rangle_0 \nonumber \\
&= \hbar^2 \! \bigg[\! \frac{\partial_i \omega_1 \partial_j \omega_1}{2}  {\rm e}^{-2i \omega_1t_{12}} \nonumber\\
 & \ \ \ + \frac{\partial_i \omega_2 \partial_j \omega_2}{2}  {\rm e}^{-2i \omega_2 t_{12}}  + \partial_i \alpha \partial_j \alpha \frac{\left( \omega_1^2 - \omega_2^2\right)^2}{4 \omega_1 \omega_2} {\rm e}^{-i (\omega_1+\omega_2)t_{12}}\bigg] \, .
\end{align}
Then, substituting this result into Eq.~(\ref{qua:metric}) and integrating, we arrive at
\begin{align}\label{lco:qclassmetric}
g^{(0)}_{ij}(x)=&\frac{\partial_i \omega_1 \partial_j \omega_1}{8 \omega_1^2} +  \frac{\partial_i \omega_2 \partial_j \omega_2}{8 \omega_2^2} \nonumber \\
&+\partial_i \alpha \partial_j \alpha \left[\frac{1}{4}\left(\frac{\omega_1}{\omega_2} 
+  \frac{\omega_2}{\omega_1} \right)-\frac{1}{2}\right]\, .
\end{align}

By using the Bohr-Sommerfeld quantization rule $I_1=I_2=\hbar/2$ and the identifications $I_1^2=I_2^2=\hbar^2$, we find that the metrics~(\ref{lco:classmetric}) and~(\ref{lco:qclassmetric}) satisfy the relation 
\begin{equation}\label{lco:relation}
g_{ij}^{(0)}(x) =\frac{1}{\hbar^{2}} g_{ij}(I;x)-\frac{1}{2} \partial_i \alpha \partial_j \alpha\,,
\end{equation}
instead of the relation~(\ref{relationmetrics}). Note that the second term in Eq.~(\ref{lco:relation}) depends on the parameters of the system, and hence in this case the classical metric~(\ref{class:metric}) does not yield the full parameter structure of the quantum metric tensor~(\ref{qua:metric}). The second term in Eq.~(\ref{lco:relation}) is a consequence of a quantum anomaly~\cite{Garcia2001}, that arises due to the ordering of the operators in the following expectation values:
\begin{equation}
\langle \hat{Q}_1 \hat{P}_2 \hat{Q}_2 \hat{P}_1 \rangle_0 + \langle \hat{Q}_2 \hat{P}_1 \hat{Q}_1 \hat{P}_2 \rangle_0 =\frac{\hbar^2}{2}\, ,
\end{equation}
whose classical counterparts are zero:
\begin{equation}
\langle Q_1 P_2 Q_2 P_1 \rangle + \langle Q_2 P_1 Q_1 P_2 \rangle = 0 \, .
\end{equation}

Now, we shall compute the classical curvature~(\ref{class:curv}).
Using the corresponding functions $\lambda_i(t)$, the average with respect to the angle variables of the non-equal-time  Poisson Brackets in Eq.~(\ref{class:curv}) gives
\begin{align}
\langle\{\lambda_{i}(t_{1}),&\lambda_{j}(t_{2})\}_{(Q_0,P_0)}\rangle \nonumber\\
&=-I_1 \cos\omega_1 t_{12} \bigg[ 4 \partial_i \omega_1 \partial_j \omega_1 \sin\omega_1 t_{12} \nonumber\\
& \ \ \ +\partial_i\alpha \partial_j\alpha\frac{(\omega_1^2-\omega_2^2)^2}{\omega_1 \omega_2}  \sin\omega_2 t_{12} \bigg] \nonumber\\
&\ \ \ -I_2 \cos\omega_2 t_{12} \bigg[ 4 \partial_i \omega_2 \partial_j \omega_2 \sin\omega_2 t_{12} \nonumber\\
&\ \ \ +\partial_i\alpha \partial_j\alpha\frac{(\omega_1^2-\omega_2^2)^2}{\omega_1 \omega_2} \sin\omega_1 t_{12} \bigg]\, .
\end{align}
Then, substituting Eq.~(\ref{lco:avg}) into Eq.~(\ref{class:curv}) and using
\begin{equation}
\intop_{-\infty}^{0}{\rm d}t_{1}\intop_{0}^{\infty} {\rm d} t_{2} \cos \omega_a t_{12} \sin \omega_b t_{12} = 0,
\end{equation}
we finally obtain that the classical curvature vanishes, $F_{ij}(I;x)=0$.  

In the quantum case, the quantum geometric tensor~(\ref{geomtesor}) is purely real and hence the Berry curvature~(\ref{qua:curvature}) is zero, $F_{ij}^{(0)}(x)=0$. Consequently, the classical and quantum curvatures are in complete agreement.

To conclude this example, we may point out that the resulting classical metric accounts for almost the entire parameter structure of the quantum metric tensor. Then here, as in the example of the linearly coupled harmonic oscillator, the parameter space of the associated classical system could in principle be employed to obtain information, or at least preliminary information, about  quantum entanglement.

\subsection{Singular Euclidean oscillator}

Here we consider the model of a singular oscillator on the two-dimensional Euclidean space, whose Hamiltonian is given by
\begin{equation}
H=\frac{\boldsymbol{p}^{2}}{2}+\frac{\alpha^{2}}{2\boldsymbol{r}^{2}}+\frac{\omega^{2}\boldsymbol{r}^{2}}{2}\, , 
\end{equation}
with $x=\{x^{i}\}=(\omega,\alpha)$ ($i,j,\dots\!=1,2$) the adiabatic parameters. This system is also known in the literature with the name of ``isotonic oscillator'', and is of special interest since it shares some properties with harmonic oscillator. Furthermore, it is has been investigated in supersymmetric quantum mechanics~\cite{CASAHORRAN1995429} and quantum rings (semiconductor ring-shaped systems)~\cite{Bellucci2011}. It should also be pointed out that the non-adiabatic Berry phase of this system was calculated in Ref.~\cite{Maamache_1996} and its non-adiabatic Hannay angle in Ref.~\cite{MAAMACHE19971}.

 In polar coordinates, this Hamiltonian reads
\begin{equation}
H=\frac{p_{r}^{2}}{2}+\frac{p_{\theta}^{2}+\alpha^{2}}{2r^{2}}+\frac{\omega^{2}r^{2}}{2},
\end{equation}
where the angular momentum $p_{\theta}$ is a constant of motion. To have orbits with a fixed energy $E$ we assume  $E>\omega\sqrt{p_{\theta}^{2}+\alpha^{2}}$. The functions ${\cal O}_{i}(t)$ for this system are
\begin{equation}\label{eq:lambda}
{\cal O}_{1}(t) =\omega r^{2}\, , \qquad
{\cal O}_{2}(t) =\frac{\alpha}{r^{2}}\, .
\end{equation}
To obtain the functions $\lambda_i (t)$,  it is convenient at this point to introduce the  action-angle variables. For the singular Euclidean oscillator, these variables are well known and read (see, for example, Ref.~\cite{Bellucci2011}):
\begin{subequations}
	\begin{align}
		&I_{\theta} =p_{\theta}\, , \\
		&I_{r} =\frac{E}{2\omega}-\frac{\tilde{p}_{\theta}}{2}\, ,\\
		&\phi_{\theta}  =\theta-\frac{p_{\theta}}{2\tilde{p}_{\theta}}\arcsin\frac{(\tilde{p}_{\theta}+\omega r^{2})\sqrt{2Er^{2}-\tilde{p}_{\theta}^{2}-\omega^{2}r^{4}}}{(E+\omega\tilde{p}_{\theta})r^{2}}\, , \\
		&\phi_{r}=-\arcsin\frac{E-\omega^{2}r^{2}}{\sqrt{E^{2}-\omega^{2}\tilde{p}_{\theta}^{2}}}\, ,\label{eq:angles}
	\end{align}
\end{subequations}
where $\tilde{p}_{\theta}=\sqrt{p_{\theta}^{2}+\alpha^{2}}$. Notice that the energy in terms of action variables can be written as
\begin{equation}
E=\omega\left(2I_{r}+\sqrt{I_{\theta}^{2}+\alpha^{2}}\right)\, . \label{eq:energy}
\end{equation} 

Now, solving Eq.~(\ref{eq:angles}) for $r$ and writing $\phi_{r}=\omega_{r}t-\beta$
where $\omega_{r}=\partial E/\partial I_{r}=2\omega$ and $\beta$
is a constant, we have
\begin{equation}
r(t)  =\frac{\left[E+\sqrt{E^{2}-\omega^{2}\tilde{p}_{\theta}^{2}}\sin(2\omega t-\beta)\right]^{1/2}}{\omega}\, , \label{eq:rfunctime}
\end{equation}
from which it follows that 
\begin{equation}
p_{r}(t)  =\dot{r} =\frac{\sqrt{E^{2}-\omega^{2}\tilde{p}_{\theta}^{2}}\cos(2\omega t-\beta)}{\left[E+\sqrt{E^{2}-\omega^{2}\tilde{p}_{\theta}^{2}}\sin(2\omega t-\beta)\right]^{1/2}}\, .\label{eq:functime}
\end{equation}
It remains to express $\beta$ in terms of the initial conditions. By setting $r_{0}=r(t=0)$ and $p_{r0}=p_{r}(t=0)$, we get
\begin{equation}
\cos\beta=\frac{\omega r_{0}p_{r0}}{\sqrt{E^{2}-\omega^{2}\tilde{p}_{\theta}^{2}}}\, ,\qquad \sin\beta=\frac{E-\omega^{2}r_{0}^{2}}{\sqrt{E^{2}-\omega^{2}\tilde{p}_{\theta}^{2}}}\, .\label{eq:beta}
\end{equation}
Thus, substituting (\ref{eq:rfunctime}) with (\ref{eq:beta}) into (\ref{eq:lambda}), we find the functions $\lambda_i(t)$:
\begin{subequations}
	\begin{align}
		\lambda_{1}(t) & =\frac{E+\omega r_{0}p_{r0}\sin2\omega t-(E-\omega^{2}r_{0}^{2})\cos2\omega t}{\omega}\, , \label{eq:lambd2}\\
		\lambda_{2}(t) & =\frac{\alpha\omega^{2}}{E+\omega r_{0}p_{r0}\sin2\omega t-(E-\omega^{2}r_{0}^{2})\cos2\omega t}\, .\label{eq:lambd}
	\end{align}
\end{subequations}
Furthermore, the initial conditions, $r_{0}$ and $p_{r0}$, in terms of initial action-angle variables, $\phi_{r0}=\phi_{r}(t=0)$ and $I_r(\equiv I_r(t=0))$, are
\begin{subequations}
	\begin{align}
		&r_{0}=\frac{1}{\omega}\left(E+\sqrt{E^{2}-\omega^{2}\tilde{p}_{\theta}^{2}}\sin\phi_{r0}\right)^{1/2}\, ,\\
		&p_{r0}=\frac{\sqrt{E^{2}-\omega^{2}\tilde{p}_{\theta}^{2}}\cos\phi_{r0}}{\left(E+\sqrt{E^{2}-\omega^{2}\tilde{p}_{\theta}^{2}}\sin\phi_{r0}\right)^{1/2}}\, ,
	\end{align}
\end{subequations}
where $\tilde{p}_{\theta}=\sqrt{I_{\theta}^{2}+\alpha^{2}}$ and $E$ is given by Eq.~(\ref{eq:energy}). Then, using these expressions, Eqs.~(\ref{eq:lambd2}) and (\ref{eq:lambd}) become
\begin{subequations}
	\begin{align}
		\lambda_{1}(t) & =\frac{E}{\omega}\left[1+a\sin(\phi_{r0}+2\omega t)\right]\, , \label{eq:lambdas-2}\\
		\lambda_{2}(t) & =\frac{\alpha\omega^{2}}{E\left[1+a\sin(\phi_{r0}+2\omega t)\right]}\, ,\label{eq:lambdas-1}
	\end{align}
\end{subequations}
where $a=\sqrt{1-\omega^{2}\tilde{p}_{\theta}^{2}/E^{2}}$ with $0<a<1$.

We are now in a position to compute the classical metric~(\ref{class:metric}). Using Eqs.~(\ref{eq:lambdas-2}) and (\ref{eq:lambdas-1}) and defining $\Lambda_{ij}:=\langle\lambda_{i}(t_{1})\lambda_{j}(t_{2})\rangle -\langle\lambda_{i}(t_{1})\rangle \langle\lambda_{j}(t_2)\rangle$ where  $\langle f\rangle=\frac{1}{2\pi}\intop_{0}^{2\pi}d\phi_{r0} f$, we find
\begin{subequations}
	\begin{align}
		&\Lambda_{11}=\frac{(E^{2}-\omega^{2}\tilde{p}_{\theta}^{2})}{2\omega^{2}}\cos2\omega t_{12}\, , \label{eq:L11}\\
		&\Lambda_{12}=-\frac{\alpha}{\tilde{p}_{\theta}}\left(E-\omega\tilde{p}_{\theta}\right)\cos2\omega t_{12}\, , \label{eq:L12} \\
		&\Lambda_{22}= \frac{\alpha^{2}\omega^{3}}{E\tilde{p}_{\theta}(1-a^{2}\cos^{2}\omega t_{12})}-\frac{\alpha^{2}\omega^{2}}{\tilde{p}_{\theta}^{2}}\, , \label{eq:L22}
	\end{align}
\end{subequations}
with $t_{12}=t_1-t_2$. The components $g_{11}$ and $g_{12}$ of the classical metric are then obtained by substituting Eqs.~(\ref{eq:L11}) and (\ref{eq:L12}), respectively, into Eq.~(\ref{class:metric}), and using (\ref{gho:avgcos}). By doing so, we obtain 
\begin{align}
	&g_{11}(I;x)  =\frac{I_{r}^{2}+I_{r}\sqrt{I_{\theta}^{2}+\alpha^{2}}}{2\omega^{2}}\, ,\label{eq:g11}\\
	&g_{12}(I;x) =\frac{-\alpha I_{r}}{2\omega\sqrt{I_{\theta}^{2}+\alpha^{2}}}\, .\label{eq:g12}
\end{align}

The component $g_{22}$ follows by substituting Eq.~(\ref{eq:L22}) into Eq.~(\ref{class:metric}). After some algebra, we get
\begin{align}
	g_{22}(I;x) =\frac{\alpha^{2}\omega^{2}}{\tilde{p}_{\theta}^{2}}\intop_{-\infty}^{0}{\rm d}t_{1}\intop_{0}^{\infty}{\rm d}t_{2}\,\left[1-\frac{\sqrt{1-a^{2}}}{1-a^{2}\cos^{2}\omega t_{12}}\right].\label{eq:g22}
\end{align}
To deal with this integral, let us consider the Fourier expansion of the function $f(T)\equiv\frac{1}{1-a^{2}\cos^{2}\omega T}$,
\begin{equation} \label{eq:fourier}
f(T)=\frac{c_{0}}{2}+\sum_{n=1}^{\infty}c_{n}\cos 2n\omega T\, ,
\end{equation}
where 
\begin{equation}
c_{n}=\frac{2\omega}{\pi}\intop_{0}^{\pi/\omega} {\rm d}T\,\frac{\cos2n\omega T}{1-a^{2}\cos^{2}\omega T}\, .
\end{equation}
By plugging~(\ref{eq:fourier}) with $T\equiv t_{12}$ into Eq.~(\ref{eq:g22}) and using Eq.~(\ref{gho:avgcos}), we arrive at
\begin{align}
	g_{22}(I;x) & =-\frac{\alpha^{2}\omega^{2}}{\tilde{p}_{\theta}^{2}}\sqrt{1-a^{2}}\sum_{n=1}^{\infty}c_{n}\intop_{-\infty}^{0}{\rm d}t_{1}\intop_{0}^{\infty}{\rm d}t_{2}\,\cos2n\omega t_{12}\nonumber \\
	&=\frac{\alpha^{2}\omega^{2}}{2\pi\tilde{p}_{\theta}E}\intop_{0}^{\pi/\omega}{\rm d}T \frac{1}{1-a^{2}\cos^{2}\omega T} \sum_{n=1}^{\infty}\frac{\cos2n\omega T}{n^{2}} \nonumber\\
	&=\frac{\alpha^{2}\omega^{2}}{2\tilde{p}_{\theta}E}\intop_{0}^{\pi/\omega}{\rm d}T \frac{\left(\frac{\pi}{6}-\omega T+\frac{\omega^{2}T^{2}}{\pi}\right)}{1-a^{2}\cos^{2}\omega T}\, , \label{eq:g22a}
\end{align}
where in the last line we have used the series representation of the quadratic Bernoulli polynomial, $B_{2}(z)=z^{2}-z+\frac{1}{6}=\frac{1}{\pi^{2}}\sum_{n=1}^{\infty}\frac{\cos2n\pi z}{n^{2}},\quad 0\leq z\leq1$.  After solving the integrals above, the expression for the component $g_{22}$ is:
\begin{equation}
g_{22}(I;x)=\frac{\alpha^{2}}{2\tilde{p}_{\theta}^{2}}\mathrm{Li}_{2}\left(\frac{I_{r}}{I_{r}+\tilde{p}_{\theta}}\right)\, , 
\end{equation}
where $\mathrm{Li}_{2}(z)$ is the dilogarithm function. With these results at hand, it is straightforward to show that the determinant of the classical metric is
\begin{equation}
\det[g_{ij}(I;x)]=\frac{\alpha^{2}I_{r}^{2}}{4\omega^{2}\tilde{p}_{\theta}^{2}}\left[\left(1+\frac{\tilde{p}_{\theta}}{I_{r}}\right)\mathrm{Li}_{2}\left(\frac{I_{r}}{I_{r}+\tilde{p}_{\theta}}\right)-1\right]\, .
\end{equation}

We want now to compare the components of the classical metric with those coming from the quantum metric tensor. However, in this case, instead of using the Eq.~(\ref{qua:metric}), we use the following expression for the quantum metric tensor~\cite{Provost1980}:
\begin{equation}\label{eq:provost}
g_{ij}^{(0)}(x)=\mathrm{Re}\left(\langle\partial_{i}\psi_{0}|\partial_{j}\psi_{0}\rangle-\langle\partial_{i}\psi_{0}|\psi_{0}\rangle\langle\psi_{0}|\partial_{j}\psi_{0}\rangle\right)\, ,
\end{equation}
where $|\psi_{0}\rangle$ is the ground eigenstate. For the singular Euclidean oscillator, the Schr\"odinger equation is
\begin{equation}
\frac{\partial^{2}\psi}{\partial r^{2}}+\frac{1}{r}\frac{\partial\psi}{\partial r}+\frac{1}{r^{2}}\frac{\partial^{2}\psi}{\partial\theta^{2}}+\frac{2}{\hbar^{2}}\left[E-\left(\frac{\alpha^{2}}{2r^{2}}+\frac{\omega^{2}r^{2}}{2}\right)\right]\psi=0\, ,
\end{equation}
and the associated (normalized) wave function of the ground-state  is
\begin{equation}
\psi_{0}(r,\theta)=\sqrt{\frac{\omega}{\pi\alpha\Gamma\left(\frac{\alpha}{\hbar}\right)}}\left(\frac{\omega}{\hbar}\right)^{\frac{\alpha}{2\hbar}}r^{\frac{\alpha}{\hbar}}e^{-\frac{\omega}{2\hbar}r^{2}}\, ,\label{eq:wavefunction}
\end{equation}
where $\Gamma(z)$ is the gamma function. Furthermore, the ground state energy is given by $E_{0}=(\alpha+\hbar)\omega$.

Plugging Eq.~(\ref{eq:wavefunction}) into Eq.~(\ref{eq:provost}), we arrive at the following components for the quantum metric tensor:
\begin{align}
	g_{11}^{(0)}(x) & =\frac{\alpha+\hbar}{4\hbar\omega^{2}}\, , \label{eq:gq11}\\
	g_{12}^{(0)}(x) & =-\frac{1}{4\hbar\omega}\, ,\label{eq:gq12}\\
	g_{22}^{(0)}(x) & =\frac{\psi_{1}(1+\frac{\alpha}{\hbar})}{4\hbar^{2}}\, ,\label{eq:gq22}
\end{align}
where $\psi_{1}(z):=\frac{d^{2}}{dz^{2}}\ln\Gamma(z)$ is the trigamma function. Note that for any $\alpha$, this metric has a nonvanishing determinant 
\begin{equation}
\det[g_{ij}^{(0)}(x)]=\frac{\left(1+\frac{\alpha}{\hbar}\right)\psi_{1}\left(1+\frac{\alpha}{\hbar}\right)-1}{16\hbar^{2}\omega^{2}}.
\end{equation}

By comparing $g_{11}(I;x)$ with $g_{11}^{(0)}(x)$  as well as $g_{12}(I;x)$ with $g_{12}^{(0)}(x)$, it is readily seen that these components are related through~Eq.(\ref{relationmetrics}), provided that the following identifications hold: $I_{r}=\hbar/2$, $I_{r}^{2}=\hbar^{2}/2$, and $I_{\theta}=0$. Now, the expansion of $g_{22}(I;x)$ up to second order in $I_{r}$ together with the use of these identifications gives
\begin{equation}
g_{22}(I_{r},I_{\theta}=0;x)=\frac{\hbar}{4\alpha}-\frac{3\hbar^{2}}{16\alpha^{2}}\, , \label{eq:g22c}
\end{equation}
while the expansion of $\hbar^2 g_{22}^{(0)}(x)$ up to second order in $\hbar$ leads to
\begin{equation}
\hbar^{2}g_{22}^{(0)}(x)=\frac{\hbar}{4\alpha}-\frac{\hbar^{2}}{8\alpha^{2}}\, .\label{eq:gq22c}
\end{equation}
Then, by comparing Eq.~(\ref{eq:g22c}) with Eq.~(\ref{eq:gq22c}), it follows that $g_{22}(I;x) $ and $g_{22}^{(0)}(x)$ satisfy the relation (\ref{relationmetrics}) (with the associated identifications) only when we retain terms to first order in $\hbar$. The discrepancy in terms of order two (or higher) in $\hbar$ may be a consequence of the fact that the quantum metric tensor takes into account loop integrals, whereas the classical metric does not. It is also worth comparing the determinants of these metrics.  Using the previous identifications and Eq.~(\ref{eq:g22c}), the determinant of the classical metric correct to third order in $\hbar$ is $\det[g_{ij}(I;x)]=\hbar^3/ 64 \alpha \omega^2$, whereas the analogous expression for quantum metric is $\hbar^4\det[g_{ij}^{(0)}(x)]=\hbar^3/ 32 \alpha \omega^2$. This shows that both determinants agree up to the second order in $\hbar$.

We conclude this example by computing the classical and quantum curvatures. The expression (\ref{class:curv}) for the classical curvature requires the evaluation of the non-equal-time Poisson bracket $\{\lambda_{1}(t_{1}),\lambda_{2}(t_{2})\}_{(r_{0},p_{r0})}$.  However, on account of the canonical invariance of the bracket, we can use $\phi_{r0}$ and $I_r$ instead of $r_{0}$ and $p_{r0}$. Taking advantage of this and using (\ref{eq:lambdas-2}) and (\ref{eq:lambdas-1}), we have 
\begin{align}
	\{\lambda_{1}&(t_{1}),\lambda_{2}(t_{2})\}_{\phi_{r0},I_r}\nonumber \\
	&=\frac{\partial\lambda_{1}(t_{1})}{\partial\phi_{r0}}\frac{\partial\lambda_{2}(t_{2})}{\partial I_{r}}-\frac{\partial\lambda_{1}(t_{1})}{\partial I_{r}}\frac{\partial\lambda_{2}(t_{2})}{\partial\phi_{r0}} \nonumber \\
	&=\frac{4\alpha\omega^{2}\sin\omega t_{12}}{E}\left\{ \frac{\cos\omega t_{12}+a\sin[\omega (t_{1}+t_{2})+\phi_{r0}]}{\left[1+a\sin(2\omega t_{2}+\phi_{r0})\right]^{2}}\right\}\, . \label{eq:pb}
\end{align}
Here, we omitted the derivatives with respect to $\phi_{\theta0}=\phi_{\theta}(0)$, since $\lambda_{i}(t)$ do not depend on it. The average of Eq.~(\ref{eq:pb}) yields
\begin{equation}
\langle\{\lambda_{1}(t_{1}),\lambda_{2}(t_{2})\}_{\phi_{r0},I_r}\rangle=\frac{2\alpha\omega}{\tilde{p}_{\theta}}\sin2\omega t_{12}\, .
\end{equation}
Thus, plugging this expression into Eq.~(\ref{class:curv}) and bearing in mind
Eq.~(\ref{gho:avgsin}), we find that the classical curvature vanishes, $F_{12}(I,x)=0$. On the quantum side, by writing the Berry curvature as
\begin{equation}
F^{(0)}_{ij}(x)=-{\rm Im}  \left(\langle\partial_{i}\psi_{0}|\partial_{j}\psi_{0}\rangle- \langle\partial_{j}\psi_{0}|\partial_{i}\psi_{0}\rangle \right)\, ,
\end{equation}
and recalling that  $\psi_{0}(r,\theta)$ is a real wave function, it is straightforward to see that $F^{(0)}_{12}(x)=0$. Therefore, the classical and quantum curvatures are in complete agreement.

\subsection{Spin-half in a magnetic field}

In our final example, we would like to show how our approach can also be applied to deal with a classical system corresponding to a quantum system with fermions. To this end, we choose the system of a particle with  spin $1/2$ in a magnetic field $\boldsymbol{B}=(B_{1},B_{2},B_{3})$. The Hamiltonian reads
\begin{equation}\label{spin:quaH}
\hat{H}=\frac{1}{2} \mu \boldsymbol{\sigma} \cdot \boldsymbol{B} \, ,
\end{equation}
where $\boldsymbol{\sigma}=(\sigma_1,\sigma_2,\sigma_3)$ is a three-vector of Pauli matrices and $x=\{x_{i}\}=(B_{1},B_{2},B_{3})$ ($i,j,\dots\!=1,2,3$) are adiabatic parameters. This system was used by Berry~\cite{Berry45} with the purpose of illustrating the existence of the geometrical phase, while its classical counterpart was addressed by Gozzi \textit{et al.}~\cite{Gozzi2388_1987,Gozzi2752_1990}, who calculated the corresponding Hannay angle.  Recently, experimental 	measurements of the quantum metric tensor and the full quantum geometric tensor in two-level systems described by Eq.~(\ref{spin:quaH}) were reported in Refs.~\cite{Tan2019} and~\cite{2018arXiv181112840Y}, respectively (see also \cite{Ozawa2018,2019arXiv190103219G}).

Following Ref.~\cite{Gozzi2752_1990}, we introduce two  complex Grassmann variables $\{\psi_{a}\}$ ($a,b,\dots\!=1,2$) satisfying $\psi_{a}\psi_{b}+\psi_{b}\psi_{a}=0$, and write the Hamiltonian of the classical associated model as
\begin{equation}\label{spin:classH}
H=\psi^{\dagger}M(\boldsymbol{B})\psi \, ,
\end{equation}
where $\psi=(\psi_{1},\psi_{2})^{T}$ is a Grassmann vector and 
\begin{equation}
M(\boldsymbol{B}):=\sum_{i=1}^{3}B_{i}\sigma_{i}=\begin{pmatrix}B_{3} & B_{1}-iB_{2}\\
B_{1}+iB_{2} & -B_{3}
\end{pmatrix}\, ,
\end{equation}
is a parameter-dependent Hermitian matrix. 

From Eq.~(\ref{spin:classH}), it is direct to see that the functions ${\cal O}_{i}(t)$ are given by
\begin{equation}\label{spin:Oi}
{\cal O}_{i}(t)=\left(\frac{\partial H}{\partial B_{i}}\right)_{\psi^{\dagger},\psi}=\psi^{\dagger}(t)\sigma_{i} \psi(t)\, .
\end{equation}
Next, we need to obtain the functions $\lambda_i (t)$. However, to do this, it is necessary first to write Eq.~(\ref{spin:Oi}) in terms of normal modes $\tilde{\psi}$. Let us consider the transformation
\begin{equation}
\psi=U\tilde{\psi}\, ,
\end{equation} 
where $U$ is a parameter-dependent  unitary matrix given by
\begin{equation}
U=\begin{pmatrix}\frac{B_{1}-iB_{2}}{\sqrt{2B(B-B_{3})}} & \frac{B_{1}-iB_{2}}{\sqrt{2B(B+B_{3})}}\\
\frac{\sqrt{B-B_{3}}}{\sqrt{2B}} & -\frac{\sqrt{B+B_{3}}}{\sqrt{2B}}
\end{pmatrix}\, ,
\end{equation}
and such that
\begin{equation}
\tilde{M}=U^{\dagger}MU=\begin{pmatrix}\Omega_{1} & 0\\
0 & \Omega_{2}
\end{pmatrix}=\begin{pmatrix}B & 0\\
0 & -B
\end{pmatrix}\, ,
\end{equation}
with $B=(B_{1}^{2}+B_{2}^{2}+B_{3}^{2})^{1/2}$. In terms of the new variables $\tilde{\psi}_a$ and their corresponding  momenta $\tilde{\Pi}_{a}=i\tilde{\psi}_{a}^{\ast}$, the Hamiltonian is diagonalized as  $H=\Omega_{1}\tilde{\psi}_{1}^{\ast}\tilde{\psi}_{1}+\Omega_{2}\tilde{\psi}_{2}^{\ast}\tilde{\psi}_{2}$ and the functions ${\cal O}_{i}(t)$ take the form
\begin{equation}\label{spin:Oi2}
{\cal O}_{i}(t)=\tilde{\psi}^{\dagger}(t)\tilde{\sigma}_{i}\tilde{\psi}(t),
\end{equation}
where we have defined the matrices $\tilde{\sigma}_{i} := U^{\dagger}\sigma_{i}U$, which can be written as
\begin{equation}
\begin{split}\label{spin:sigmatilde}
	&\tilde{\sigma}_{1}=\begin{pmatrix}\frac{B_{1}}{B} & -\frac{B_{1}B_{3}+iB_{2}B}{B\sqrt{B_{1}^{2}+B_{2}^{2}}}\\
		-\frac{B_{1}B_{3}-iB_{2}B}{B\sqrt{B_{1}^{2}+B_{2}^{2}}} & -\frac{B_{1}}{B}
	\end{pmatrix}\, ,\\
	&\tilde{\sigma}_{2}=\begin{pmatrix}\frac{B_{2}}{B} & -\frac{B_{2}B_{3}-iB_{1}B}{B\sqrt{B_{1}^{2}+B_{2}^{2}}}\\
		-\frac{B_{2}B_{3}+iB_{1}B}{B\sqrt{B_{1}^{2}+B_{2}^{2}}} & -\frac{B_{2}}{B}
	\end{pmatrix}\, ,\\
	&\tilde{\sigma}_{3}=\begin{pmatrix}\frac{B_{3}}{B} & \frac{\sqrt{B_{1}^{2}+B_{2}^{2}}}{B}\\
		\frac{\sqrt{B_{1}^{2}+B_{2}^{2}}}{B} & -\frac{B_{3}}{B}
	\end{pmatrix}\, .
\end{split}
\end{equation}
Now, defining the Poisson brackets with respect to the variables $(\tilde{\psi}_a,\tilde{\Pi}_{a})$ as
\begin{eqnarray}
\{f,g\}_{(\tilde{\psi},\tilde{\psi}^{\dagger})}=i\sum_{a=1}^{2}\left(f\frac{\overset{\leftarrow}{\partial}}{\partial\tilde{\psi}_{a}^{\ast}}\frac{\overset{\rightarrow}{\partial}}{\partial\tilde{\psi}_{a}}g+f\frac{\overset{\leftarrow}{\partial}}{\partial\tilde{\psi}_{a}}\frac{\overset{\rightarrow}{\partial}}{\partial\tilde{\psi}_{a}^{\ast}}g\right)\, , \nonumber \\
\end{eqnarray}
where $\overset{\leftarrow}{\partial}$ and $\overset{\rightarrow}{\partial}$ denote  the right  and left derivatives, the equations of motion turn out to be $\dot{\tilde{\psi}}_{a}=\{H,\tilde{\psi}_{a}\}_{(\tilde{\psi},\tilde{\psi}^{\dagger})}=-i\Omega_{a}\tilde{\psi}_{a}$ and $\dot{\tilde{\psi}}_{a}^{\ast}=\{H,\tilde{\psi}_{a}^{\ast}\}_{(\tilde{\psi},\tilde{\psi}^{\dagger})}=i\Omega_{a}\tilde{\psi}_{a}^{\ast}$, and have the solutions
\begin{equation}
\tilde{\psi}_{a}(t)=\tilde{\psi}_{a0}{\rm e}^{-i\Omega_{a}t},\qquad \tilde{\psi}_{a}^{\ast}(t)=\tilde{\psi}_{a0}^{\ast}{\rm e}^{i\Omega_{a}t}\, ,
\end{equation}
where $\tilde{\psi}_{a0}=\tilde{\psi}_{a}(t=0)$. Consequently, the action-angle variables  $(\phi_{a},I_{a})$ are given by
\begin{equation}
I_{a}=\frac{1}{2\pi}\oint\mathrm{d}t\tilde{\Pi}_{a}\dot{\tilde{\psi}}_{a}=\tilde{\psi}_{a0}^{\ast}\tilde{\psi}_{a0}, \ \ \phi_{a}=\Omega_{a}t+\phi_{a0} \, ,
\end{equation}
with  $\phi_{a0}=\phi_{a}(t=0)$,  which allows us to cast the Hamiltonian into the form $H=\Omega_{1}I_{1}+\Omega_{2}I_{2}$. Hence, the normal modes in terms of initial variables and time read
\begin{equation}
\tilde{\psi}_{a}(t)=\tilde{\psi}_{a0}{\rm e}^{-i\phi_{a0}}{\rm e}^{-i\Omega_{a}t}\, .\label{eq:soltime}
\end{equation}
Inserting Eq.~(\ref{eq:soltime}) into Eq.~(\ref{spin:Oi2}), we arrive at
\begin{equation}\label{spin:lambda}
\lambda_{i}(t)=\sum_{a,b}\tilde{\psi}_{a0}^{\ast}\tilde{\psi}_{b0}{\rm e}^{i(\Omega_{a}-\Omega_{b})t}{\rm e}^{i(\phi_{a0}-\phi_{b0})}\tilde{\sigma}_{iab}\, ,
\end{equation}
where  $\tilde{\sigma}_{iab}$ are the components of the matrices (\ref{spin:sigmatilde}).

Now we have all the ingredients we need to compute the classical metric. Considering that the only dependence on $\phi_{a0}$ in Eq.~(\ref{spin:lambda}) is through ${\rm e}^{i(\phi_{a0}-\phi_{b0})}$ and  using  $\langle f\rangle=\frac{1}{(2\pi)^2}\intop_{0}^{2\pi}{\rm d}\phi_{10}\intop_{0}^{2\pi}{\rm d}\phi_{20}\,f$ for the average over the fast variables, we obtain
\begin{align}\label{spin:avg}
\langle\lambda_{i}&(t_{1})\lambda_{j}(t_{2})\rangle-\langle\lambda_{i}(t_{1})\rangle\langle\lambda_{j}(t_{2})\rangle \nonumber\\
&=-I_{1}I_{2}\left({\rm e}^{2iB t_{12}}\tilde{\sigma}_{i12}\tilde{\sigma}_{j21}+{\rm e}^{-2iBt_{12}}\tilde{\sigma}_{i21}\tilde{\sigma}_{j12}\right),
\end{align}
where $t_{12}=t_1-t_2$, and we have used $\langle e^{i(\phi_{a0}-\phi_{b0})}\rangle=\delta_{ab}$ and $\langle e^{i(\phi_{a0}-\phi_{b0}+\phi_{c0}-\phi_{d0})}\rangle=\delta_{ab}\delta_{cd}+\delta_{a1}\delta_{b2}\delta_{c2}\delta_{d1}+\delta_{a2}\delta_{b1}\delta_{c1}\delta_{d2}$. Then, plugging Eq.~(\ref{spin:avg}) into Eq.~(\ref{class:metric}) and integrating, we arrive at the classical metric 
 \begin{equation}
 g_{ij}(I;x)=-\frac{I_{1}I_{2}}{4B^{2}}\left(\tilde{\sigma}_{i12}\tilde{\sigma}_{j21}+\tilde{\sigma}_{i21}\tilde{\sigma}_{j12}\right)\, ,
 \end{equation}
which can be written in matrix form  as
\begin{equation}
g_{ij}(I;x)=-\frac{I_{1}I_{2}}{2B^{4}}\begin{pmatrix}B_{2}^{2}+B_{3}^{2} & -B_{1}B_{2} & -B_{1}B_{3}\\
-B_{1}B_{2} & B_{1}^{2}+B_{3}^{2} & -B_{2}B_{3}\\
-B_{1}B_{3} & -B_{2}B_{3} & B_{1}^{2}+B_{2}^{2}
\end{pmatrix}\, .\label{spin:cmetric}
\end{equation}
Notice that this metric can be simplified considerably if we perform a  simple change of coordinates. Writing the magnetic field  in spherical coordinates as $(B_1,B_2,B_3)=B(\sin\theta \cos\varphi,\sin\theta \sin\varphi,\cos\theta)$, choosing $y=\{y^{i}\}=(B,\theta,\varphi)$ as the adiabatic parameters, and recalling that $g_{ij}(I;x)$ transforms as a tensor~\cite{GGVmetric2019}, this classical metric takes the form
\begin{equation}\label{spin:cmetric2}
{g'}_{ij}(I;y)=-\frac{I_{1}I_{2}}{2}\begin{pmatrix}0 & 0 & 0\\
0 & 1 & 0\\
0 & 0 & \sin^{2}\theta
\end{pmatrix}.
\end{equation}

As it is customary, at this stage we compare the classical metric with its quantum counterpart. In the quantum setting, a spin $1/2$ particle is a system described by a two-dimensional Hilbert space spanned by the eigenstates with spin projections $+1/2$ and $-1/2$, which means that in this case we have two quantum metric tensors, namely ${g'}^{(+)}_{ij}(y)$ and ${g'}^{(-)}_{ij}(y)$. Such metrics are well--known in the literature~\cite{SHI-JIAN2010,chruscinski2012geometric}, and are given by
\begin{equation}\label{spin:qmetric}
{g'}^{(+)}_{ij}(y)={g'}^{(-)}_{ij}(y)=\frac{1}{4}\begin{pmatrix}0 & 0 & 0\\
0 & 1 & 0\\
0 & 0 & \sin^{2}\theta
\end{pmatrix}\, ,
\end{equation}
where $y=\{y^{i}\}=(B,\theta,\varphi)$ are the adiabatic parameters.

 By comparing Eqs.~(\ref{spin:cmetric2}) and (\ref{spin:qmetric}), we can see that the metrics are related as follows:
\begin{equation}
 {g'}_{ij}^{(\pm)}(y)=-\frac{1}{2I_{1}I_{2}}{g'}_{ij}(I;x)\, .
\end{equation}
Therefore, the whole parameter structure of the quantum metric tensors is captured by the classical metric, modulo an appropriate quantization rule for the action variables. 

On the other hand, it is worth noting that these metrics have vanishing determinants and rank two, which implies that one of the parameters is redundant and can be treated as a constant.  If, for instance, instead of taking $y=\{y^{i}\}=(B,\theta,\varphi)$ as the adiabatic parameters, we say that $z=\{z^{i'}\}=(\theta,\varphi)$ ($i',j',\dots\!=1,2$) are the adiabatic parameters and consider that $B$ is a nonvanishing constant, then resulting classical metric is 
\begin{equation}\label{spin:cmetric3}
g_{i'j'}(I;z)=-\frac{I_{1}I_{2}}{2}\begin{pmatrix}
 1 & 0\\
 0 & \sin^{2}\theta
\end{pmatrix}\, ,
\end{equation}
which has a nonzero determinant. By the way,  $g_{i'j'}(I;z)$ is--up to a minus sign--the standard metric on the surface of a sphere with radius  $R=(I_{1}I_{2}/2)^{1/2}$.

We now turn to the calculation of the classical curvature.  The non-equal-time Poisson brackets between $\lambda_{i}(t_{1})$ and $\lambda_{j}(t_{2})$, with respect to the initial variables $(\tilde{\psi}_{a0},\tilde{\Pi}_{a0})$, yields
\begin{align}\label{spin:pb}
&\{\lambda_{i}(t_{1}),\lambda_{j}(t_{2})\}_{\psi_0,\psi^{\dagger}_0}\nonumber\\
&=i\sum_{a}\left(\!\lambda_{i}(t_{1})\frac{\overset{\leftarrow}{\partial}}{\partial\tilde{\psi}_{a0}^{\ast}}\frac{\overset{\rightarrow}{\partial}}{\partial\tilde{\psi}_{a0}}\lambda_{j}(t_{2}) \!+\!\lambda_{i}(t_{1})\frac{\overset{\leftarrow}{\partial}}{\partial\tilde{\psi}_{a0}}\frac{\overset{\rightarrow}{\partial}}{\partial\tilde{\psi}_{a0}^{\ast}}\lambda_{j}(t_{2})\right) \nonumber\\
&=i\sum_{a,b,c}\tilde{\psi}_{b0}^{\ast}\tilde{\psi}_{c0} \biggl\{ -{\rm e}^{i\left[(\Omega_{a}-\Omega_{c})t_{1}+(\Omega_{b}-\Omega_{a})t_{2}\right]}\tilde{\sigma}_{iac}\tilde{\sigma}_{jba}\nonumber\\
&\qquad \qquad \qquad+{\rm e}^{i\left[(\Omega_{b}-\Omega_{a})t_{1}+(\Omega_{a}-\Omega_{c})t_{2}\right]}\tilde{\sigma}_{iba}\tilde{\sigma}_{jac} \biggr\} {\rm e}^{i(\phi_{b0}-\phi_{c0})}\, .
\end{align}
Taking the average over the fast variables of Eq.~(\ref{spin:pb}), we obtain
\begin{align}\label{spin:pbavg}
	&\langle\{\lambda_{i}(t_{1}),\lambda_{j}(t_{2})\}_{\psi_0,\psi^{\dagger}_0}\rangle \nonumber\\
	 & =i\sum_{a,b}I_{b}\left[-{\rm e}^{i(\Omega_{a}-\Omega_{b})t_{12}}\tilde{\sigma}_{iab}\tilde{\sigma}_{jba}+{\rm e}^{-i(\Omega_{a}-\Omega_{b})t_{12}}\tilde{\sigma}_{iba}\tilde{\sigma}_{jab}\right]\nonumber \\
	& =i(I_{1}-I_{2})\left[{\rm e}^{i(\Omega_{1}-\Omega_{2})t_{12}}\tilde{\sigma}_{i12}\tilde{\sigma}_{j21}-{\rm e}^{-i(\Omega_{1}-\Omega_{2})t_{12}}\tilde{\sigma}_{i21}\tilde{\sigma}_{j12}\right]\, .
\end{align}
Finally, plugging Eq.~(\ref{spin:pbavg}) into Eq.~(\ref{class:curv}) and integrating, we arrive at the classical curvature
\begin{equation}
F_{ij}(I;x)=\frac{i(I_{1}-I_{2})}{4B^{2}}\left(\tilde{\sigma}_{i21}\tilde{\sigma}_{j12}-\tilde{\sigma}_{i12}\tilde{\sigma}_{j21}\right) \, ,
\end{equation}
which can be written as 
\begin{align}
&F_{12}(I;x)=\frac{(I_{1}-I_{2})}{2B^{3}}B_{3}\, , \qquad F_{23}(I;x)=\frac{(I_{1}-I_{2})}{2B^{3}}B_{1}\, , \nonumber\\
&F_{31}(I;x)=\frac{(I_{1}-I_{2})}{2B^{3}}B_{2} \, .
\end{align}
These components are precisely the same as those obtained in Ref.~\cite{Gozzi2388_1987}, using the standard expression for the curvature of the Hannay connection. It is worth noting that, in spherical coordinates $y=\{y^{i}\}=(B,\theta,\varphi)$, the only
nonvanishing component is
\begin{equation}\label{spin:classF}
F_{23}(I;y)=\frac{(I_{1}-I_{2})}{2}\sin\theta.
\end{equation}

In the quantum setting, the system has two Berry curvatures, $F^{(+)}_{i,j}(y)$ and $F^{(-)}_{i,j}(y)$, which are also well--known~\cite{Berry45,chruscinski2012geometric}, and have the non-zero components
\begin{equation}\label{spin:quanF}
F^{(+)}_{23}(y)=-F^{(-)}_{23}(y)=-\sin\theta\, .
\end{equation}
By comparing Eqs.~(\ref{spin:classF}) and~(\ref{spin:quanF}), it is straightforward to see that the relationship between the classical and quantum curvatures is
\begin{equation}
F^{(+)}_{ij}(y)=-F^{(-)}_{ij}(y)=-
\frac{2}{(I_{1}-I_{2})}F_{ij}(I;y).
\end{equation}

In this way, we have seen in this example that our expressions for the classical metric and curvature in the parameters space, namely Eqs.~(\ref{class:metric}) and (\ref{class:curv}), can be applied not only to typical integrable systems but also to Grassmannian models corresponding to quantum systems with fermions, leading to the expected results. It should be pointed out that this is the first time that the classical analog of the quantum metric tensor has been determined for a system involving fermions. In this sense, this example serves as a warm-up for more realistic systems. 

\section{Conclusions}\label{Conclusions}

We have introduced new expressions for the curvature of the Hannay connection and the metric proposed in Ref.~\cite{GGVmetric2019}, which are the classical counterparts of the Berry curvature and the quantum metric tensor, respectively. Furthermore, we have established the semiclassical relation between the quantum metric tensor and the classical metric proposed in Ref.~\cite{GGVmetric2019}. The new expressions, as well as the semiclassical relation, were obtained by performing a semiclassical approximation of the quantum metric tensor and the Berry curvature  in the Lagrangian formalism. A distinguishing feature of our approach to obtain the classical metric is that it can be applied to a wide variety of systems, even those whose quantum counterpart involves fermions, and without prior knowledge of any generating function. 

We have shown the applicability and validity of our approach in five different systems. For the generalized harmonic oscillator, we have seen that the approach yields the already known expressions for the classical analogs of the Berry curvature and the quantum metric tensor and that the semiclassical relation between the classical and quantum metrics is satisfied. In the case of the symmetric coupled harmonic oscillators, the resulting classical metric and curvature have precisely the same structure as their quantum counterparts, while in the case of the linearly coupled harmonic oscillators, the classic and quantum metrics differ by one term, which is a consequence of a quantum anomaly that arises from the ordering of the operators. Regarding the singular Euclidean oscillator, our analysis has shown that the classical and quantum metrics agree to first order in $\hbar$, which is somehow expected since the classical metric only has tree--level contributions and the quantum metric involves loop corrections.   This result is remarkable since the singular Euclidean oscillator involves a potential that cannot be treated in a perturbative way using standard procedures.  Finally, for the system of a particle with  spin $1/2$ in a magnetic field, we have shown how our approach can be extended to deal with Grassmannian models corresponding to quantum systems with fermions. In this case, the classical curvature obtained from our approach  is the same as that found by Gozzi \textit{et al.}~\cite{Gozzi2752_1990} using the standard expression, and the classical metric turns out to have the same structure as its quantum counterpart.  As has been mentioned, in all the systems that were presented, in order to satisfy Eq.~(\ref{relationmetrics}), we needed to adjust the quantization rule for different powers of the action variables to some numerical coefficients. This can be thought to be analogous to what happens in Ref.~\cite{Henneaux1994}, where the algebraic `classical' procedure to find all the possible anomalies in Yang-Mills theory, matches the quantum result only up to numerical coefficients.

It is important to point out that we have not found in the literature any study on the Berry curvature, the quantum metric tensor, or their classical counterparts for the cases of the symmetric coupled harmonic oscillators, the linearly coupled harmonic oscillators, and the singular Euclidean oscillator. The same follows for the classical  metric of  the system of a spin-half in a magnetic field. Thus, the expressions that we have obtained here for these classical and quantum geometrical structures are new and can be used to further investigate the nature of the parameter spaces of  the corresponding classical and quantum systems.  For instance, it is interesting to note that the classical metric associated to the symmetric coupled harmonic oscillators can be split as the sum of two parts, each of which involves only one action variable of the  uncoupled subsystems. This separability property is not present in the case of the linearly coupled harmonic oscillators, which may suggest that this system could be more `entangled' than the symmetric coupled harmonic oscillators.

In summary, for all the systems analyzed in this paper, the resulting parameter structure of the classical metrics and curvatures are the same or almost the same as those of their corresponding quantum counterparts. Therefore, our study of the parameter space of a classical integrable system, which is purely classical since it only involves averages over classical (angle) variables and does not require any prior knowledge of the quantum system Hilbert space, reveals that this space captures all, or at least a good part of, the information that can be extracted from the parameter space of the associated quantum system. This means, for instance, that the parameter space  obtained in the classical setting may be used to gain a first insight into quantum phase transitions or quantum entanglement. This shows the importance of  this space, and that it is worth studying.

In this sense, it will be very interesting to apply our approach to other more realistic systems such as those of condensed matter physics where the Berry curvature and the quantum metric tensor play an important role, in particular, toward understanding the existence of tensor monopoles in the parameter space~\cite{Palumbo2018}. In this line of work, it would be also interesting to analyze, from the point of view of the parameter space  in the classical setting, the effect of a quantum dissipative environment on a fermionic system~\cite{Henriet2018}.   Furthermore, in the context of many-body systems, the quantum metric tensor can be related to the mean-square fluctuation of the macroscopic bulk polarization in insulators~\cite{Souza2000}. In this line of work, it would be interesting to extend our classical metric to many-body fermionic systems using Grassmann variables and see whether it can predict similar results to those of the quantum case. In the same spirit, we could try to analyze a Bose-Einstein condensate in the mean-field approximation using the Gross-Pitaevskii equation. There exists a mixed quantum-classical approach where one can resort to a type of action-angle variables and compute both Hannay angle and Berry phase~\cite{Niu2017}. However, in purely classical terms, the Inverse Scattering Transform (IST) is known to provide the interpretation that some nonlinear partial differential equations are Hamiltonian systems where the IST can be thought of as a canonical transformation to action-angle variables~\cite{Ablowitz1991}. This suggests that we could use our procedure, once it has been properly generalized to an infinite number of degrees of freedom, to compute our classical metric and see whether it contains information regarding the quantum aspects of the condensate. Finally, it would also be worth exploiting the extension of our approach to the classical analog of the non-Abelian quantum metric tensor~\cite{YuQuan2010}, and classical systems with chaotic dynamics~\cite{Robbins631}. 

\begin{acknowledgements}
J.D.V.  would like to thank  Prof. A. Balatsky  for his kind hospitality at Dynamic Quantum Matter and the invitation to participate in this special issue in Annalen der Physik.  This work was partially supported by DGAPA-PAPIIT Grants No. IN103716 and No. IN103919, CONACyT project 237503. J. Alvarez is supported by CONACyT Ph.D. scholarship (No.419420). D. Guti\'errez-Ruiz is supported with a CONACyT Ph.D. scholarship (No. 332577). D. Gonzalez is supported with a DGAPA-UNAM postdoctoral fellowship. 
\end{acknowledgements}

\bibliography{references}

\end{document}